\documentclass[prx,longbibliography,nofootinbib,notitlepage,10pt]{revtex4-1}
\usepackage[utf8]{inputenc} 
\usepackage[T1]{fontenc}    
\usepackage[colorlinks]{hyperref}       
\usepackage{url}            
\usepackage{booktabs}       
\usepackage{amsfonts}       
\usepackage{nicefrac}       
\usepackage{microtype}      
\usepackage{amsmath}
\usepackage{mathrsfs}
\usepackage[pdftex]{graphicx}
\usepackage{tikz}
\usepackage{placeins}
\usepackage{bbm}
\usepackage[normalem]{ulem}
\usepackage{enumerate}

\usepackage{float}
\usepackage{amsmath,amssymb,setspace}
\usepackage{subfigure}
\usepackage{color}
\usepackage{braket}
\usepackage{algorithm}
\usepackage{algpseudocode}
\newcommand\bea{\begin{eqnarray}}
\newcommand\eea{\end{eqnarray}}

\algdef{SE}[DOWHILE]{Do}{doWhile}{\algorithmicdo}[1]{\algorithmicwhile\ #1}

\usepackage{mathabx}

\makeatletter
\def\l@subsection#1#2{}
\def\l@subsubsection#1#2{}
\makeatother

\begin{document}

\title{Towards the Practical Application of Near-Term Quantum Computers in\\ Quantum Chemistry Simulations: A Problem Decomposition Approach}
\date{\today}
\author{Takeshi Yamazaki}
\email{takeshi.yamazaki@1qbit.com}
\affiliation{1QB Information Technologies (1QBit), 458-550 Burrard Street, Vancouver, BC, Canada, V6C 2B5}

\author{Shunji Matsuura}
\email{shunji.matsuura@1qbit.com}
\affiliation{1QB Information Technologies (1QBit), 458-550 Burrard Street, Vancouver, BC, Canada, V6C 2B5}

\author{Ali Narimani}
\email{narimanyali@gmail.com}
\affiliation{1QB Information Technologies (1QBit), 458-550 Burrard Street, Vancouver, BC, Canada, V6C 2B5}

\author{Anushervon Saidmuradov}
\email{anush.saidmuradov@1qbit.com}
\affiliation{1QB Information Technologies (1QBit), 458-550 Burrard Street, Vancouver, BC, Canada, V6C 2B5}

\author{Arman Zaribafiyan}
\email{arman.zaribafiyan@1qbit.com}
\affiliation{1QB Information Technologies (1QBit), 458-550 Burrard Street, Vancouver, BC, Canada, V6C 2B5}

\begin{abstract}

With the aim of establishing a framework to efficiently perform the practical application of quantum chemistry simulation on near-term quantum devices, we envision a hybrid quantum--classical framework for leveraging problem decomposition (PD) techniques in quantum chemistry. Specifically, we use PD techniques to decompose a target molecular system into smaller subsystems requiring fewer computational resources. In our framework, there are two levels of hybridization. At the first level, we use a classical algorithm to decompose a target molecule into subsystems, and utilize a quantum algorithm to simulate the quantum nature of the subsystems. The second level is in the quantum algorithm. We consider the quantum--classical variational algorithm that iterates between an expectation estimation using a quantum device and a parameter optimization using a classical device.
We investigate three popular PD techniques for our hybrid approach: the fragment molecular-orbital (FMO) method, the divide-and-conquer (DC) technique, and the density matrix embedding theory (DMET). We examine the efficacy of these techniques in correctly differentiating conformations of simple alkane molecules. In particular, we consider the ratio between the number of qubits for PD and that of the full system; the mean absolute deviation; and the Pearson correlation coefficient and Spearman's rank correlation coefficient. Sampling error is introduced when expectation values are measured on the quantum device. Therefore, we study how this error affects the predictive performance of PD techniques. The present study is our first step to opening up the possibility of using quantum chemistry simulations at a scale close to the size of molecules relevant to industry on near-term quantum hardware.

\end{abstract}

\maketitle

\section*{Introduction}

Accurate modelling of chemical processes requires a highly accurate description of their quantum nature. In general, however, simulating quantum systems on classical computers is  a computationally demanding task.
The dimension of the Hilbert space in quantum systems increases exponentially with respect to the system size, and without any approximations it becomes intractable to diagonalize a Hamiltonian or even to store the full state vector for a small system of about five atoms~\cite{Head-Gordon-2008-58}.

Recently, there has been increasing interest in a new paradigm of computation, namely, quantum computing.
The idea of applying quantum devices to simulate quantum systems dates back to Richard Feynman's 1982 proposal~\cite{Feynman:1982}. An advantage of using a quantum device for simulating quantum systems is that the required computational resources scale only polynomially with the size of the system. Based on this observation, a quantum algorithm for simulating locally interacting fermionic quantum systems, a phase estimation algorithm (PEA), was used~\cite{AbramsLloyd:PEA1997, Aspuru-Guzik:2005aa}.
This opened up the possibility of conducting studies in quantum chemistry on quantum devices, and has been used to estimate molecular energies in various experiments~\cite{LanyonQCex, PhysRevLett.104.030502, Wang2014, PhysRevLett.118.100503}.

While PEA predicts accurate energies, it will be difficult to use it for larger molecules on near-term quantum devices due to the fact that a deep circuit is required. A variational method, called the variational quantum eigensolver (VQE), was proposed~\cite{Peruzzo:2014aa, VQE2, VQE3}. This method requires a shallower circuit compared to PEA and has the advantage of mitigating systematic errors in quantum devices by using the degrees of freedom introduced by its variational parameters.

\begin{figure}
    \centering
    \includegraphics[width=9cm]{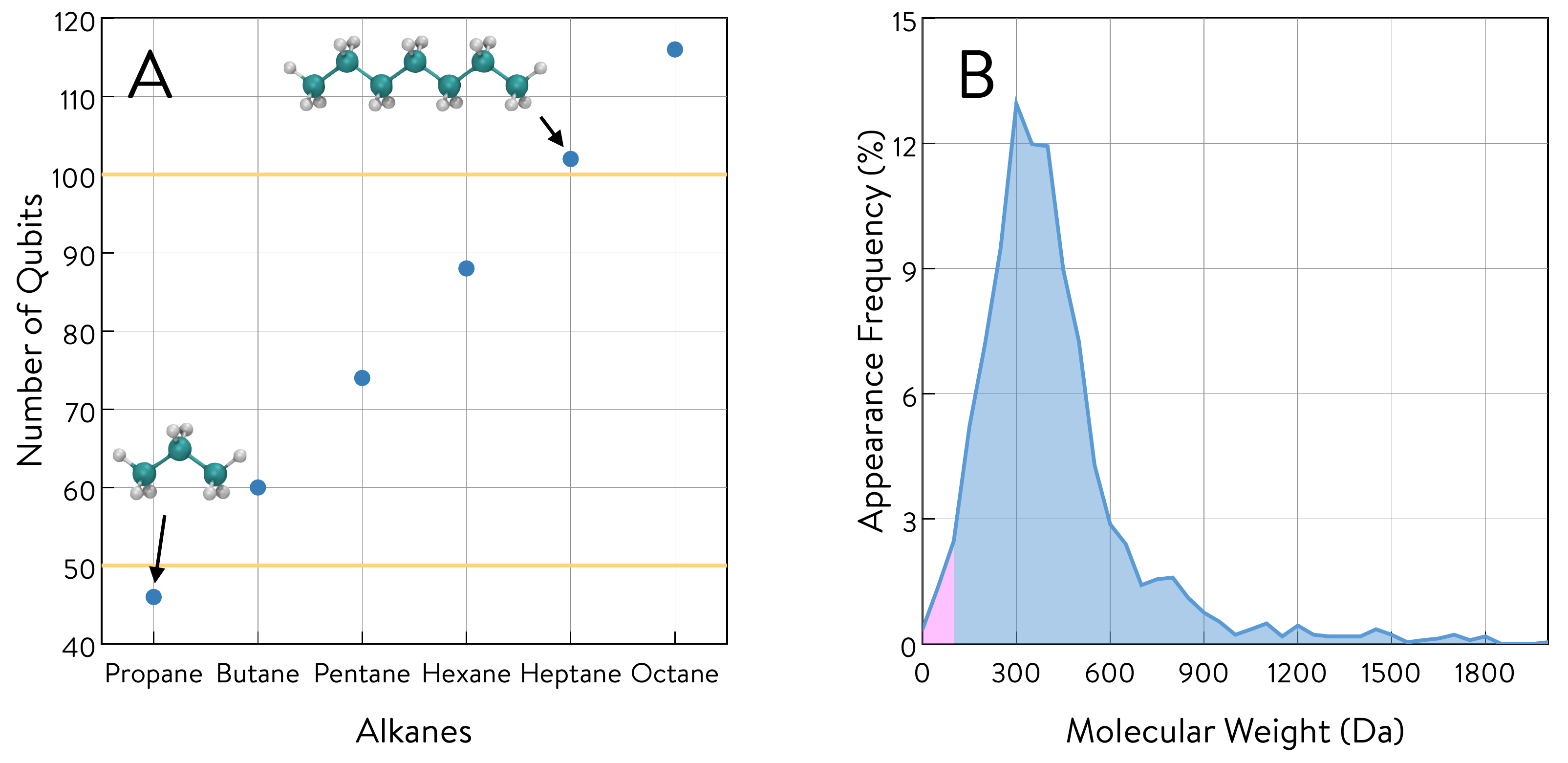}
   \caption{(A) The number of qubits required to simulate alkanes using the minimal basis. (B) The appearance frequency of approved drug molecules as a function of molecular weight. The area coloured in pink represents the molecules that weigh less than 100 Da.}
   \label{fig:alkanes_MW}
\end{figure}

The power of quantum computing depends on various factors, such as the coherence time, the gate fidelities, and the gate operation times. The gate depth is one of the most important metrics, and there have been various studies for efficient state preparations by shortening the gate depth in the context of quantum chemistry simulation~\cite{1506.05135, PhysRevA.92.042303, 1706.00023, Kivlichan-2017-1711.04789}.
Another important metric is the number of qubits.
In the second quantization picture of the molecular Hamiltonian,
the Hilbert space is spanned by spin-orbital creation and annihilation operators.
Under Jordan--Wigner or Bravyi--Kitaev~\cite{BK_trans_2, BK_trans_1} transformations, electron states are mapped into qubit states where the number of spin orbitals and the number of qubits are the same.
If we assume that a near-term quantum device will be equipped with 50--100 qubits, such a device would be able to accommodate molecules ranging from propane (three carbon atoms) to heptane (seven carbon atoms) if we consider a family of simple alkane molecules with a minimal basis set, as shown in Fig.~\ref{fig:alkanes_MW}A. We note that it has been shown that one can reduce the number of qubits required to simulate a molecule by considering its symmetries~\cite{IBM_symmetry, Alan-Martinis:PhysRevX.2017}.

From a practical perspective, the molecular size we can simulate on such near-term devices will still be rather small for the purpose of materials design. For example, the molecular weight of heptane is 100 Da, and if we consider the approved drug molecules in DrugBank 5.0~\cite{Wishart:2018aa}, the ratio of molecules that weigh less than 100 Da is approximately 4\%, as shown in Fig.~\ref{fig:alkanes_MW}B. Even with the rapid progress in the development of quantum devices, it is unlikely that we will be able to simulate the remaining 96\% of the molecules directly on near-term quantum devices. It should also be noted that the above resource estimation is based on the minimal basis set, and, for a reliable prediction of energy, it is common practice to consider larger basis sets~\cite{Szabo-1996}. In such cases, the molecular size we can actually simulate will become even smaller.

Therefore, we are in need of an alternative strategy to simulate large molecules on quantum devices. Towards this end, in this paper we introduce and assess problem decomposition (PD) techniques that are developed in the area of quantum chemistry on \emph{classical} hardware for quantum chemistry simulation on near-term \emph{quantum} hardware. PD techniques are computational techniques for decomposing the target molecular system into smaller subsystems that require fewer computational resources, estimating the electronic structure of each subsystem, and then combining them to obtain the electronic structure of the entire system. 

One of the first PD techniques to be employed in quantum chemistry is the divide-and-conquer (DC) technique proposed by Yang in 1991~\cite{Kobayashi:2011aa}. Since then, a number of PD techniques have been proposed, such as the elongation method~\cite{Imamura:1991aa}, the fragment molecular-orbital (FMO) method~\cite{Kitaura:1999aa}, the molecular fractionation with conjugate caps approach~\cite{Zhang:2003aa}, and generalized X-Pol theory~\cite{Gao:2010aa}. We refer the interested reader to the recent review about large-scale computations in chemistry~\cite{Akimov:2015aa}.

In this study, our main interest is in the efficiency of PD techniques for near-term quantum devices, in terms of the number of qubits required and in their applicability to studying molecules of a size relevant to industry. Although there have been previous studies focusing on large-scale quantum chemistry simulations on small-scale quantum computational resources using the dynamical mean-field theory (DMFT)~\cite{Bauer-2016-1510.03859} and the density matrix embedding theory (DMET)~\cite{Rubin-2016-1610.06910}, to our knowledge, this is the first study to assess PD techniques for near-term quantum hardware from the perspective of practical applications, including the differentiation of conformers.

We note that the quantum gate depth, the other important metric of quantum hardware performance, is polynomially proportional to the size of molecule we simulate. Therefore, we can expect that the PD techniques not only decrease the number of qubits required, but make the gate depth shallower by decomposing the molecule into smaller subsystems. In addition, we note that it is possible to incorporate PD techniques into any other ideas for reducing the circuit depth, such as the use of a shallower state preparation ansatz~\cite{1506.05135, PhysRevA.92.042303, 1706.00023, Aspuru-Guzik:2005aa, Wecker:2014aa, Peruzzo:2014aa, Kandala-2017-1704.05018, Babbush-2017-1706.00023, Kivlichan-2017-1711.04789} in order to realize more-efficient quantum chemistry simulations on near-term quantum hardware.\\

This paper is organized as follows. In Sec.~\ref{sec: framework}, we introduce a framework for using PD in a hybrid architecture. We then further explain the hybrid quantum--classical VQE algorithm used to estimate the electronic structure of the resulting subsystems. In Sec.~\ref{sec: methods}, we describe the context of conformational comparisons and the metrics we used to assess PD techniques together with the molecular systems that we used. We conclude this section by providing a more detailed description of three PD methods we considered, as well as a brief explanation about computational tools and our setup. In Sec.~\ref{sec: results}, we present the results. In Sec.~\ref{sec:sampling}, we estimate sampling errors and error tolerance of our computational results. Sec.~\ref{sec: summary} concludes the paper by providing a summary of results and possible future work.

\section{Quantum Chemistry Simulation on Near-Term Quantum Devices}
\label{sec: framework}
Near-term quantum computing hardware is characterized as noisy intermediate-scale quantum (NISQ) hardware. The primary reason behind this characterization is the absence of error correction. The number of qubits is too limited to enable error correction, while the coherence time is not sufficiently long and/or the gate fidelities are too low to allow for deep quantum circuits. Therefore, applications are limited to algorithms that are executable with shallow circuits. The importance of reducing the resources required for quantum computation is not limited to NISQ hardware. Even with digital universal quantum computers, fault-tolerant quantum resources will remain more scarce and expensive compared to classical resources. Thus, it is critical to think of quantum computers as co-processors for classical computers in a hybrid computing architecture. We call this quantum--classical hybrid computing. In order to maximize the advantage of using hybrid architectures, identifying how to decompose and distribute computing tasks across different hardware technologies becomes essential. Therefore, studying the impact of PD is important not only from a resource-reduction perspective, but is indispensable for the practical and scalable applications of future digital quantum computing hardware.

In this section, we introduce a high-level picture for a quantum--classical hybrid computing platform in the context of quantum chemistry simulations, and describe how variational algorithms and, in particular, the VQE algorithm fits into this picture. In order to design a hybrid architecture, it is important to assess how the PD techniques are able to efficiently decompose and accurately represent the original problem. Therefore, we propose a test application for assessing and comparing the effectiveness of these approaches.

\subsection{The proposed quantum--classical hybrid framework}
PD has always been an advantageous strategy for solving large, complex problems with limited computational resources. Not only does it break down the problem into components which can be less complex than the original problem, it makes it easy to parallelize the computational effort. The idea of using PD to establish a more efficient use of quantum computing resources in a hybrid architecture of quantum and classical devices was proposed earlier in the context of solving classical combinatorial optimization problems as well~\cite{NRZ2017}.

\begin{figure}[htbp]
\includegraphics[width=9cm]{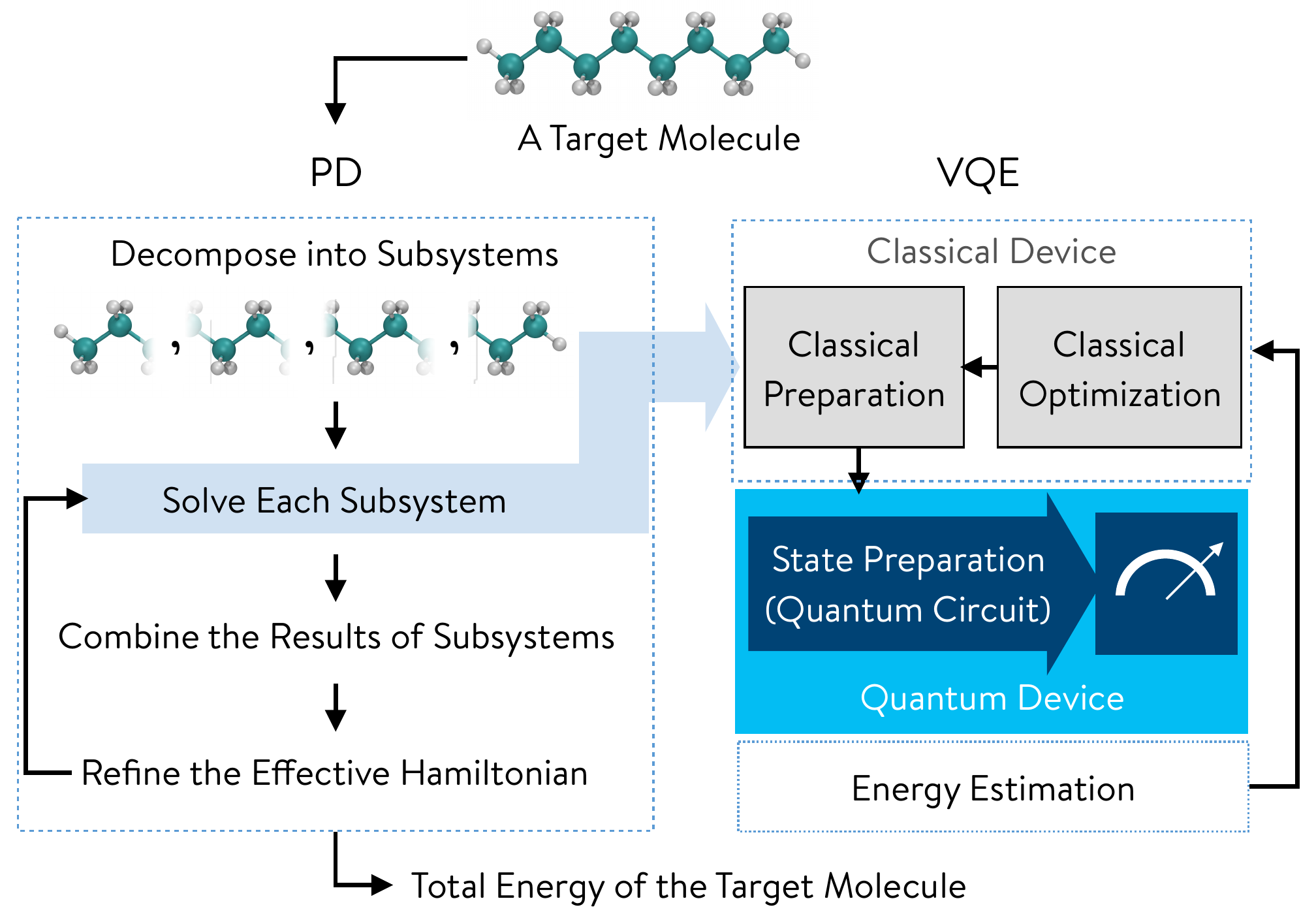}
\caption{A conceptual and schematic illustration of the hybrid framework of leveraging PD in quantum chemistry simulation. The reader should note that there are two levels of hybridization in this approach: the PD of the target molecule is performed on a classical computing device, and the more complex problem of simulating each subproblem is handled by the quantum algorithm. In this particular example, the quantum algorithm itself is a hybrid quantum--classical variational algorithm that iterates between expectation estimation using a quantum device and parameter optimization using a classical device. The PD technique dictates how different components of this framework are implemented and incorporated. 
Depending on the PD technique, the electronic structure of subsystem is solved iteratively by refining the effective Hamiltonian.
}
\label{fig:framework}
\end{figure}

We propose a general hybrid quantum--classical framework for the simulation of large chemical compounds. The specifics of this framework depend on the chosen PD technique. As we will see in Sec.~\ref{sec:PDtechniques}, different PD techniques break down the problem based on different concepts and, consequently, they have different approaches for aggregating the solution of each component to the solution for the original problem. Despite these differences, a common concept among all PD techniques is the use of a classical algorithm to break the computational task into subproblems. This is conceptually illustrated using smaller pieces of the original molecule in Fig.~\ref{fig:framework}. Then, the simulation of each subproblem is addressed using a quantum approach. Depending on the structure of the PD technique used, it may or may not be possible to parallelize the simulation of a plurality of subproblems. After each subproblem has been simulated, the results are combined on the classical computer and compiled to recreate a simulation of the large molecule. 

As mentioned earlier, there exist different quantum algorithms for estimating the total energy of a molecule (or, in this case, a fragment of a molecule). Recently, there has been a lot of interest in understanding variational algorithms, thanks to the fact that they require a shallower circuit for execution on quantum hardware, hence their applicability to near-term quantum devices. Using the same argument, we use the VQE algorithm for the purpose of this study. However, a certain class of PD ideas is also easily compatible and able to be integrated with the use of a PEA.

The VQE algorithm is a hybrid method wherein both quantum and classical computing resources are used to perform an estimation of total energy (see Fig.~\ref{fig:framework}).
The basic steps in the VQE algorithm are as follows. We start with an initial state such as the Hartree--Fock state~$|\psi_0\rangle$.
A quantum device makes a unitary transformation $U(\vec{t})$ characterized by variational parameters $\vec{t}_{i}$ on the initial state:
$|\psi(\vec{t}_{i})\rangle=U(\vec{t}_{i})|\psi_0\rangle$.
There are various possibilities for this unitary transformation $U(\vec{t})$, such as a unitary coupled cluster~\cite{Peruzzo:2014aa, VQE2, VQE3} 
and the low-depth gate ansatz~\cite{LDCA}.
We then measure the energy $E_i=\langle\psi(\vec{t}_{i})| \hat{H} |\psi(\vec{t}_{i})\rangle$,
where $\hat{H}$ is a molecular Hamiltonian represented using qubits. The energy value and the variational parameters $\vec{t}_{i}$
are sent to a classical device where a classical optimizer returns a set of updated variational parameters $\vec{t}_{i+1}$ that could provide a lower energy than $E_i$.
A quantum device then generates a new quantum state with the new variational parameters $|\psi(\vec{t}_{i+1})\rangle=U(\vec{t}_{i+1})|\psi_0\rangle$.
We repeat this process until the energy converges.
The energies of various  
small molecules such as H$_2$, LiH, and BeH$_2$ have been computed on quantum devices using the VQE algorithm~\cite{Alan-Martinis:PhysRevX.2017, Kandala-2017-1704.05018}.

\section{Methodology and Experiment Setup}
\label{sec: methods}

\subsection{Metrics for assessing problem decomposition techniques}
When we try to simulate large molecules, there is another challenge outside of the high computational cost, that is, the degree of conformational freedom.
In general, the degree of conformational freedom increases superlinearly as the number of atoms increases, and many of them lie within a very small energy window. For example, in the case of proteins, globally different structures of the same protein lie within a few $k_{\text{B}}T$ of each other~\cite{Onuchic:2004aa}. Therefore, for practical applications such as materials design, it is essential to distinguish the different conformers and select the appropriate conformers on which to work. Hence, a PD method for large molecules has to have not only the features to reduce the computational resources required and to predict a reasonably accurate total energy, but also the capability to distinguish different conformers. 
With this in mind, in the present study, we generate samples of conformers of a target molecule and then perform the energy calculation both with and without using PD techniques for each conformer to create a scatter plot between these two energies. Then, we assess the PD methods based on: 1) the ratio between the number of qubits for PD and that for the full system; 2) the mean absolute deviation (MAD) between energies with and without a PD technique; and 3) the Pearson correlation coefficient $\rho_{\text{P}}$ and Spearman's rank correlation coefficient $\rho_{\text{S}}$ between energies with and without using a PD technique.

We note that the following assessments are all based on classical quantum chemistry calculations on classical computers, as the ideal case for a noiseless quantum computer. The number of qubits estimation is done within the framework of second quantization picture.\\

\subsection{Molecular systems examined}
\begin{figure}[h]
\includegraphics[width=9cm]{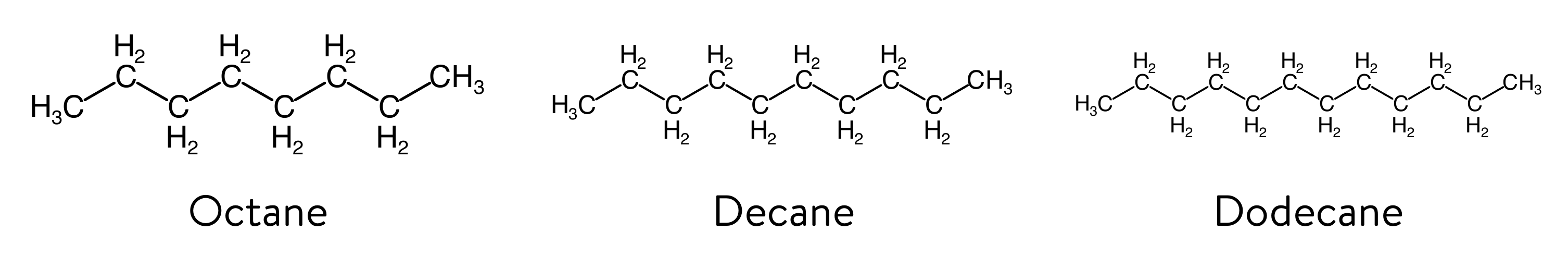}
\caption{Three straight-chain alkanes that we examined.}
\label{fig:mols}
\end{figure}

We considered three straight-chain alkanes larger than heptane: octane (eight carbon atoms), decane (10 carbon atoms), and dodecane (12 carbon atoms) (see Fig.~\ref{fig:mols}). The molecular geometries were generated using Open~Babel~\cite{OBoyle:2011aa} (version 2.4.1) from SMILES strings and their conformers were generated with a Confab~\cite{OBoyle:2011ab} module implemented in Open~Babel using the default parameters (an RMSD cutoff of 0.5 \AA, and an energy cutoff 50 kcal/mol). The numbers of conformers generated were 52, 294, and 1694, respectively, for octane, decane, and dodecane.

\subsection{Problem decomposition methods examined}
\label{sec:PDtechniques}
We carried out total energy calculations using the coupled cluster theory with single and double excitations (CCSD) with a PD technique, and compared the result with the exact (i.e., without a PD technique) CCSD calculation. As for the PD technique, we examined the DC approach~\cite{Kobayashi:2008aa}, the FMO method~\cite{Fedorov:2005aa}, and DMET~\cite{Knizia:2012aa, Wouters:2016aa}. We provide a brief introduction to each PD technique below.

\begin{figure}[h]
\includegraphics[width=8cm]{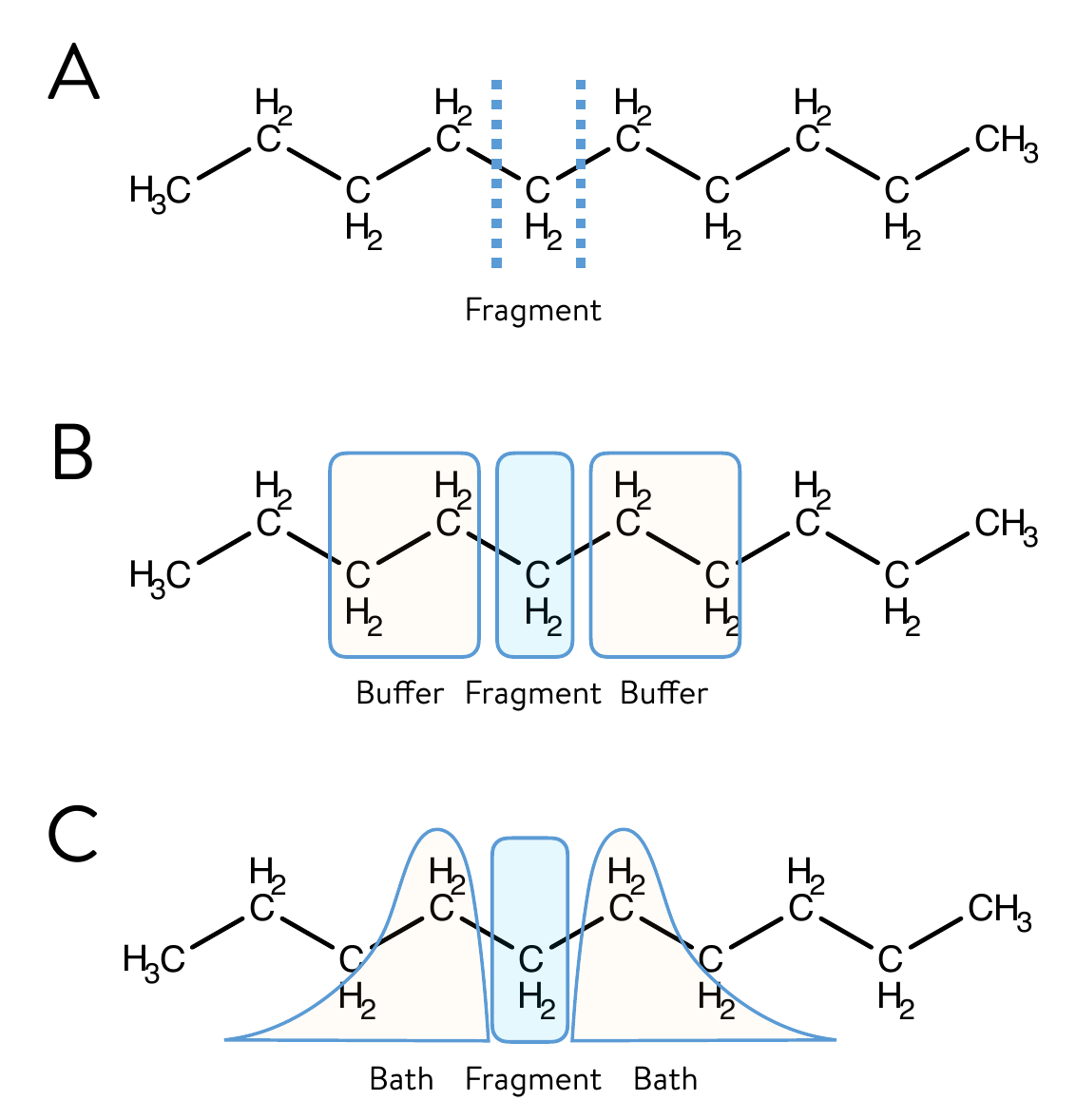}
\caption{Schematic illustration of PD techniques that we examined. (A) The FMO method, (B) the DC approach, and (C) DMET. Briefly, the FMO method, the DC approach, and DMET decompose the molecule into subsystems based on its geometry, its electron density matrix, and the entanglement between the fragment and the environment, respectively.}
\label{fig:PD_schematic}
\end{figure}

\subsubsection{The fragment molecular-orbital method}
In the FMO method, the target molecule is fragmented by detaching the bond-connecting fragments. The details of the fragmentation can be found elsewhere~\cite{Fedorov-2009}. Once the fragments have been detached from each other, fragment calculations are performed via a regular ab initio calculation of each fragment $I$ with the electrostatic potential coming from all remaining fragments. In the case of two-body FMO method, that we use in the present study, the total energy is expressed as 
\begin{equation}
E^{\text{FMO}}=\sum_{I}E_{I}+\sum_{I>J}\left[  \left( E_{IJ} - E_{I} - E_{J}  \right) \right],
\end{equation}
where $E_{I}$ is the total energy of the monomer fragment, and $E_{IJ}$ is the total energy of the dimer fragment. In order to determine the molecular orbital of each fragment, all monomer calculations are repeated until self-consistency is reached by updating the electrostatic potential, and for the dimer energy estimation, the electrostatic potential that was converged in the monomer calculation is used. The correlation energy of the total system is defined in a similar manner,
\begin{equation}
E^{\text{FMO-corr}}=\sum_{I}E^{\text{corr}}_{I}+\sum_{I>J}\left[  \left( E^{\text{corr}}_{IJ} - E^{\text{corr}}_{I} - E^{\text{corr}}_{J}  \right) \right],
\end{equation}
and the correlation calculation does not involve the self-consistency loop over the fragments, because the electrostatic potential that was converged in the monomer calculation above is used.

Using the FMO technique to estimate the correlation energy of large molecules is attractive from the viewpoint of quantum chemistry simulations on a quantum computer for the following reasons: (1) it requires the energy to be measured only once for each fragment (monomers and dimers) since the result of energy measurement of one fragment does not affect the energy measurement of the other fragments; and (2) it is applicable to both PEA and VQE algorithms since it does not require information about state vectors and depends only on the energy of the subsystems.

\subsubsection{The divide-and-conquer approach}

In the DC approach, we decompose the molecule into the subsystems composed of the fragment and the buffer region that surrounds the fragment. The density matrix of the entire system $\mathbf{D}$ is represented in terms of the density matrices of the subsystems $\alpha$ as 
\begin{equation}
D_{\mu\nu} \approx D^{\text{DC}}_{\mu\nu} =  \sum_{\alpha} D^{\alpha}_{\mu\nu}\,, \label{eq:D}
\end{equation}
where $\mu$ and $\nu$ are the atomic orbitals (bases). In the DC approximation, the local density matrix of subsystem $\alpha$ is defined as 
\begin{equation}
D_{\mu\nu}^{\alpha} = 2 p^{\alpha}_{\mu\nu} \sum_{q}^{\text{MO}(\alpha)}f_{\beta}(\epsilon_{\text{F}}-\epsilon^{\alpha}_{q})C^{\alpha}_{\mu q} C^{\alpha *}_{\nu q}\,, 
\end{equation}
where the partition matrix $\mathbf{p}^{\alpha}$ is defined as
\begin{equation}
p^{\alpha}_{\mu\nu} = 
\begin{cases}
1 &\text{for } \mu \in \text{fragment and } \nu \in \text{fragment} \\
\frac{1}{2} &\text{for } \mu \in \text{fragment and } \nu \in \text{buffer, or vice versa}\\
0 & \text{otherwise.}
\end{cases}
\end{equation}
$\epsilon^{\alpha}_{q}$ and $C^{\alpha}_{\mu q}$ are obtained by solving the Hartree--Fock equation for the subsystem $\alpha$. 
 $f_{\beta}(x)$ and $\epsilon_{\text{F}}$ are the Fermi function and the Fermi level, respectively. $\epsilon_{\text{F}}$ is globally determined after solving the Hartree--Fock equation for all subsystems to ensure that the total number of electrons in all fragments adds up to the number of electrons in the full system. Then, the Fock operator for the subsystem is reconstructed by using the local density matrix with the determined Fermi level, the calculations are iterated until convergence, and, finally, we obtain the density matrix of the total system using Eq.~\ref{eq:D}. As can be seen, the DC-based Hartree--Fock calculation provides the molecular orbital in each subsystem; therefore, it will be straightforward to obtain the correlation energy corresponding to the subsystem. However, the simple summation of the correlation energies over all subsystems cannot be considered  the correlation energy of the entire system because subsystems overlap with each other due to the existence of buffer regions. Therefore, in the DC-based correlation calculation, the correlation energy corresponding to the fragment region is extracted by rewriting the correlation energy as the sum of the atomic contributions.
 
 As in the case of the FMO technique for the correlation energy, the DC technique requires the energy to be measured only once for each subsystem, since the result of energy measurement of one subsystem does not affect the energy measurement of the other subsystems, which is beneficial for quantum chemistry simulation on quantum hardware. However, the existence of the process to rewrite the correlation energy as the sum of the atomic contributions based on the information of the density matrix will hamper the utilization of PEA, and the VQE algorithm will become the appropriate choice for the DC technique.
 
\subsubsection{The density matrix embedding theory}

In the framework of DMET~\cite{Knizia:2012aa}, we first assume that we are given the exact ground state $|\Psi\rangle$ for the system, and then perform the Schmidt decomposition of this wave function,
\begin{equation}
|\Psi\rangle = \sum_{i}^{\min(N_{A},N_{B})}\lambda_{i}|\alpha_{i}\rangle | \beta_{i} \rangle,
\end{equation}
where $|\alpha \rangle$ represents the part of the system in which we are interested, that is, the fragment, and $|\beta\rangle$ represents the rest of the system to which we refer as the ``bath'' for the fragment. $N_{A}$ and $N_{B}$ are the sizes of the Hilbert space of the fragment and the bath, respectively.
With these two states, we can define the embedding Hamiltonian as
\begin{equation}
\hat{H}^{\text{emb}} = \sum_{ijkl} |\alpha_{i} \rangle | \beta_{j} \rangle \langle \alpha_{i} | \langle  \beta_{j} | \hat{H} |\alpha_{k}\rangle|\beta_{l}\rangle \langle \alpha_{k} | \langle \beta_{l} |,
\end{equation}
which has the same ground state of the Hamiltonian as the full system $\hat{H}$. 
This is an exact result but is purely formal, because we require knowledge of the state $|\Psi\rangle$ of the full system to construct the bath states $|\beta\rangle$. In order to make this framework practical, DMET suggests that we construct the bath from an approximated state of $|\Psi\rangle$, and then use this approximated bath in a subsequent highly accurate calculation of the embedding Hamiltonian for each fragment. By convention, the exact ground state $|\Psi\rangle$ is replaced with a mean-field (i.e., Hartree--Fock) wave function. 

In DMET, the full system Hamiltonian to yield the mean-field solution is generally augmented with the correlation potential, but in the present study we use single-shot embedding~\cite{Wouters:2016aa}, where the correlation potential is set to zero. As described above, the resulting mean-field density matrix is used to define the bath states, and then used to construct the embedding Hamiltonian. Each embedding Hamiltonian yields a wave function for each fragment based on a highly accurate method (CCSD in the present study), and the total number of electrons in all local fragments is obtained as a sum of local fragment traces of the high-level, one-particle density matrix. In order to ensure that the total number of electrons in all fragments adds up to the number of electrons in the full system, the chemical potential $\mu$ is introduced,
\begin{equation}
	\hat{H}^{\text{emb}}_{I} \leftarrow \hat{H}^{\text{emb}}_{I} - \mu\sum_{r\in I} \hat{a}^{\dagger}_{r}\hat{a}_{r}\,,
\end{equation}
where $\hat{H}^{\text{emb}}_{I}$ is the embedding Hamiltonian for the fragment $I$, and $\hat{a}^{\dagger}_{r}$ and $\hat{a}_{r}$ are the creation and the annihilation operators, respectively. In the case of the single-shot embedding, the DMET calculation runs by updating the chemical potential in the embedding Hamiltonian until the total number of electrons in all fragments adds up to the number of electrons in the full system.

The necessity for a one-particle density matrix of the embedding Hamiltonian makes the VQE algorithm an appropriate quantum algorithm for the DMET calculation.

\subsection{Computational details}

GAMESS~\cite{Schmidt:1993aa} (R1 release: April, 2017) was used for the FMO and DC method calculations, and QC-DMET~\cite{Wouters-QC-DMET} code and PySCF~\cite{Sun:2018aa} (version 1.3.5) were used for the DMET calculation. For octane, 6-31G~\cite{Hehre:1972aa}, cc-pVDZ~\cite{Dunning:1989aa}, and cc-pVTZ~\cite{Dunning:1989aa} basis sets were examined, and for decane and dodecane, only 6-31G was used, based on the efficiency consideration discussed below.

In order to estimate the number of qubits required to perform the PD calculations on quantum hardware, we used the following procedures. \textbf{FMO}: The number of molecular orbitals required for the dimer calculation involving the two largest fragments (two terminal methyl groups in the present case) was doubled to obtain the number of qubits. \textbf{DC}: The number of atoms that are covered by the buffer, the sphere of 4 \AA\ radius centred at atoms in the target fragment, varies depending on the octane conformation and the position of the target fragment, and therefore the number of molecular orbitals in the DC calculation also varies. We identified the largest number of orbitals used to perform the DC calculations for all conformers, and doubled it to obtain the number of qubits. \textbf{DMET}: We followed a localization strategy that is based on the intrinsic atomic orbital construction described in Ref.~\cite{Wouters:2016aa}. We identified the largest number of correlated orbitals used in the DMET calculations, and doubled it to obtain the number of qubits.

\section{Results and Discussion}
\label{sec: results}

\subsection{Comparison between problem decomposition techniques for the octane molecule}
\begin{figure}[h]
\includegraphics[width=16cm]{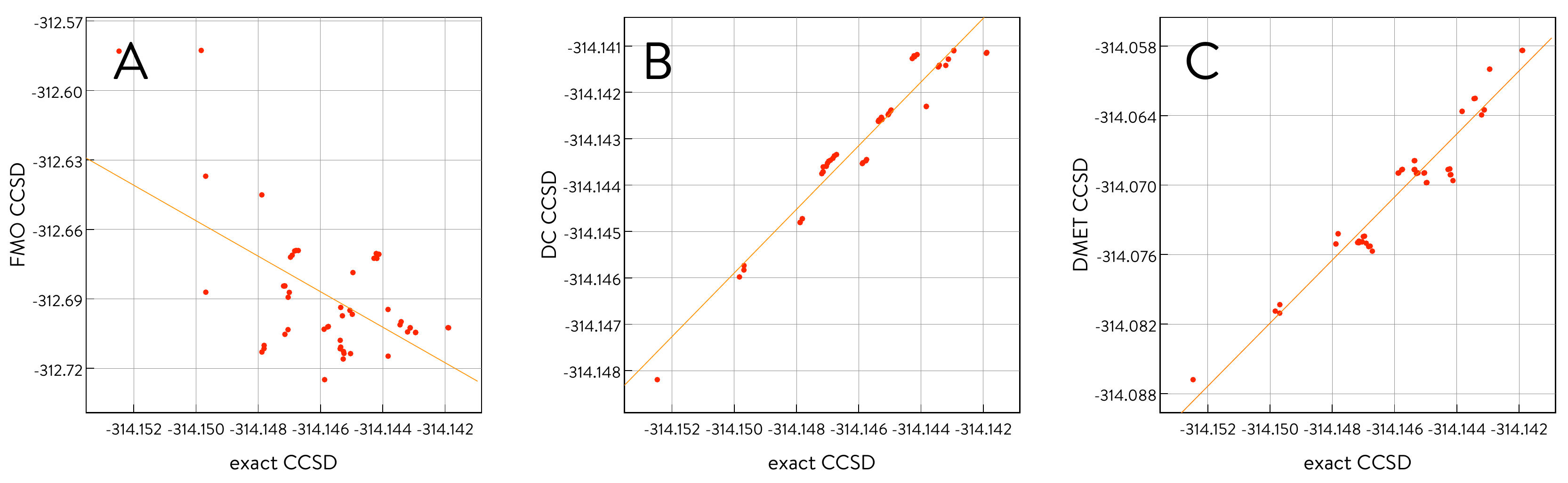}
\caption{Scatter plots of CCSD energy with the FMO method (A), the DC approach (B), and DMET (C), and the exact CCSD energy. The orange lines serve as guides for the eye.}
\label{fig:PD_comparison}
\end{figure}

Fig.~\ref{fig:PD_comparison} shows scatter plots between CCSD energy with the FMO method (A), the DC approach (B), and DMET (C), and the exact CCSD energy for 52 conformers of octane. All of the energies are in hartrees. In the present PD calculations, we considered the group consisting of one carbon atom and the hydrogen atoms bonded to it as one fragment. In order to construct the buffer region in the DC method, we used the dual-buffer, DC-based correlation scheme~\cite{Kobayashi:2008aa}, where we considered the atoms included in spheres with a 12 \AA\ radius centred at atoms in the target fragment for Hartree--Fock calculation, and a 4 \AA\ radius for the CCSD calculation. To estimate the number of qubits required, we considered the computational cost required for the correlation energy calculation. The orange line in the plots serves as a visual guide for the linear regression calculation. Table~\ref{tab:PD_comparison} summarizes the ratio (the number of qubits required to perform quantum chemistry simulations with the PD technique divided by those without the PD technique), MAD (in hartrees), $\rho_{\text{P}}$, and $\rho_{\text{S}}$ for each PD technique.

\begin{figure}[h]
\includegraphics[width=16cm]{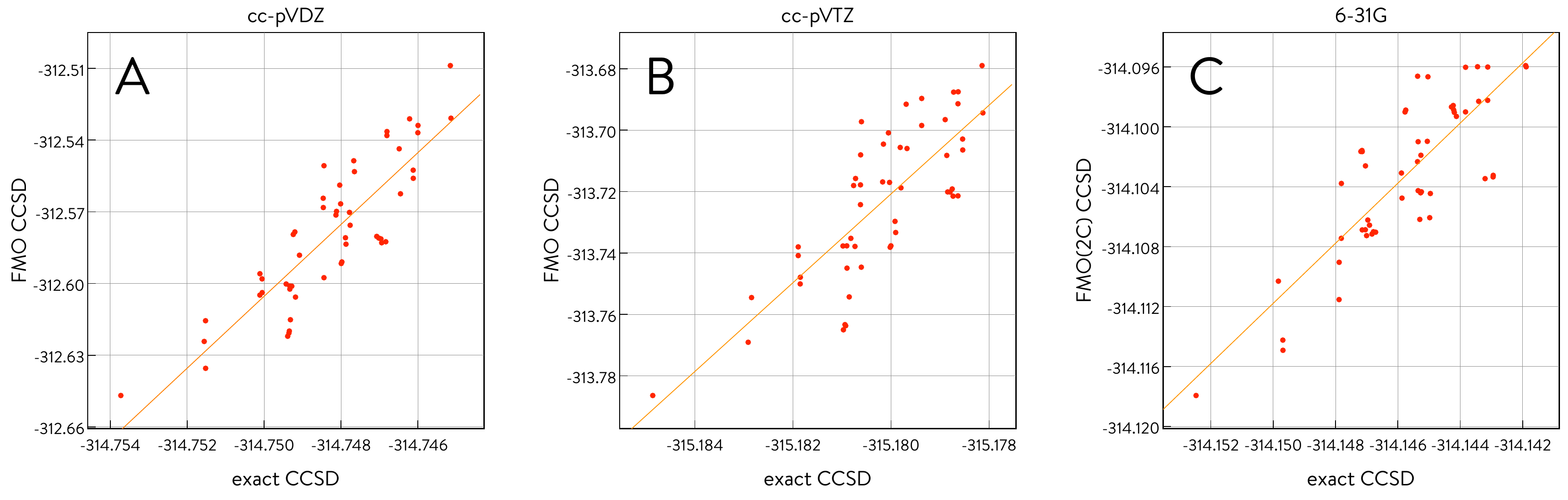}
\caption{Scatter plots of the exact CCSD energy and the CCSD energy using the FMO method with cc-pVDZ (A) and cc-pVTZ (B) basis sets. (C) Scatter plot of the exact CCSD energy and the CCSD energy using the FMO method with the 6-31G basis set and with the fragment that includes two carbon atoms and the hydrogen atoms bonded to them.}
\label{fig:FMO_with_largeBS}
\end{figure}

As can be seen, the FMO method showed a negative correlation, while the reduction in the number of qubits is significant (around a 70\% reduction), which is suitable for near-term devices with 50--100 qubits. However, we observed that the FMO method provides good correlations when a larger basis set was used (cc-pVDZ: MAD = 2.2, $\rho_{\text{P}}$ = 0.86, and $\rho_{\text{S}}$ = 0.84; cc-pVTZ: MAD=1.4, $\rho_{\text{P}}$ = 0.79, and $\rho_{\text{S}}$ = 0.77; see Fig.~\ref{fig:FMO_with_largeBS}A and B), suggesting that the negative correlation with the 6-31G basis set most likely originated from the basis set superposition error when the interaction energies between monomer fragments were estimated. Furthermore, we observed that the performance of the FMO method with the 6-31G basis set significantly improves when we use the fragment that includes two carbon atoms and the hydrogen atoms bonded to them (MAD = 0.04, $\rho_{\text{P}}$ = 0.83, and $\rho_{\text{S}}$ = 0.73), as shown in Fig.~\ref{fig:FMO_with_largeBS}C. Considering the number of qubits required to perform the calculations shown in Fig.~\ref{fig:FMO_with_largeBS}, the present observation suggests that it is desirable to have access to a quantum hardware device with more than 100 qubits in order to exploit the ability of the FMO method.

The results show that the DC technique provides reliable MAD, $\rho_{\text{P}}$, and $\rho_{\text{S}}$ values for octane; however, the number of qubits required to perform the DC correlation calculation with a 4 \AA\ buffer requires almost the same number of qubits as for the whole system. This means that the sphere with a 4 \AA\ radius centred at atoms in the target fragment covers almost the entire octane molecule, and, therefore, DC does not present a clear advantage in terms of efficiency. However, if DC can maintain its accuracy when we target the large molecular system in which the sphere with a 4 \AA\ radius becomes sufficiently small compared to the full system size, then the DC approach will be able to show good efficiency and become a good PD candidate for implementation on quantum devices. As more than 200 qubits are needed to accommodate the 4 \AA\ buffer in the case of heptane, the DC method would begin to become beneficial on quantum hardware equipped with 200--300 qubits.
Note that we tried the DC method with a smaller buffer size; however, we observed some numerical instability and, in many cases, the calculation of CCSD diverged.

Among the PD techniques examined, DMET was found to be the most reliable PD approach for octane. It reduces the qubit requirement by around 80\%, which is suitable for devices with 50--100 qubits. MAD, $\rho_{\text{P}}$, and $\rho_{\text{S}}$ were also reasonably good. Therefore, we decided to further examine DMET to see how the fragment size and the basis set size can change its efficiency (see the following subsection).


\begin{table}[h]
\caption {Comparison of PD techniques for octane. The number of qubits required was estimated based on the second quantization picture of the molecular Hamiltonian with the 6-31G basis set.} \label{tab:PD_comparison}
\setlength{\tabcolsep}{1em}
\begin{tabular}{ c | c c c }
\hline
   & FMO & DC & DMET \\
\hline
  Ratio & 60/216 & 212/216 & 46/216 \\
  MAD & 1.5 & 0.0028 & 0.075\\
  $\rho_{P}$ & -0.57 & 0.97 & 0.96 \\
  $\rho_{S}$ & -0.28 & 0.97  & 0.86 \\
  \hline
\end{tabular}
\end{table}


\subsection{Basis set size and fragment size dependency of the density matrix embedding theory}

\begin{figure}[h]
\includegraphics[width=16cm]{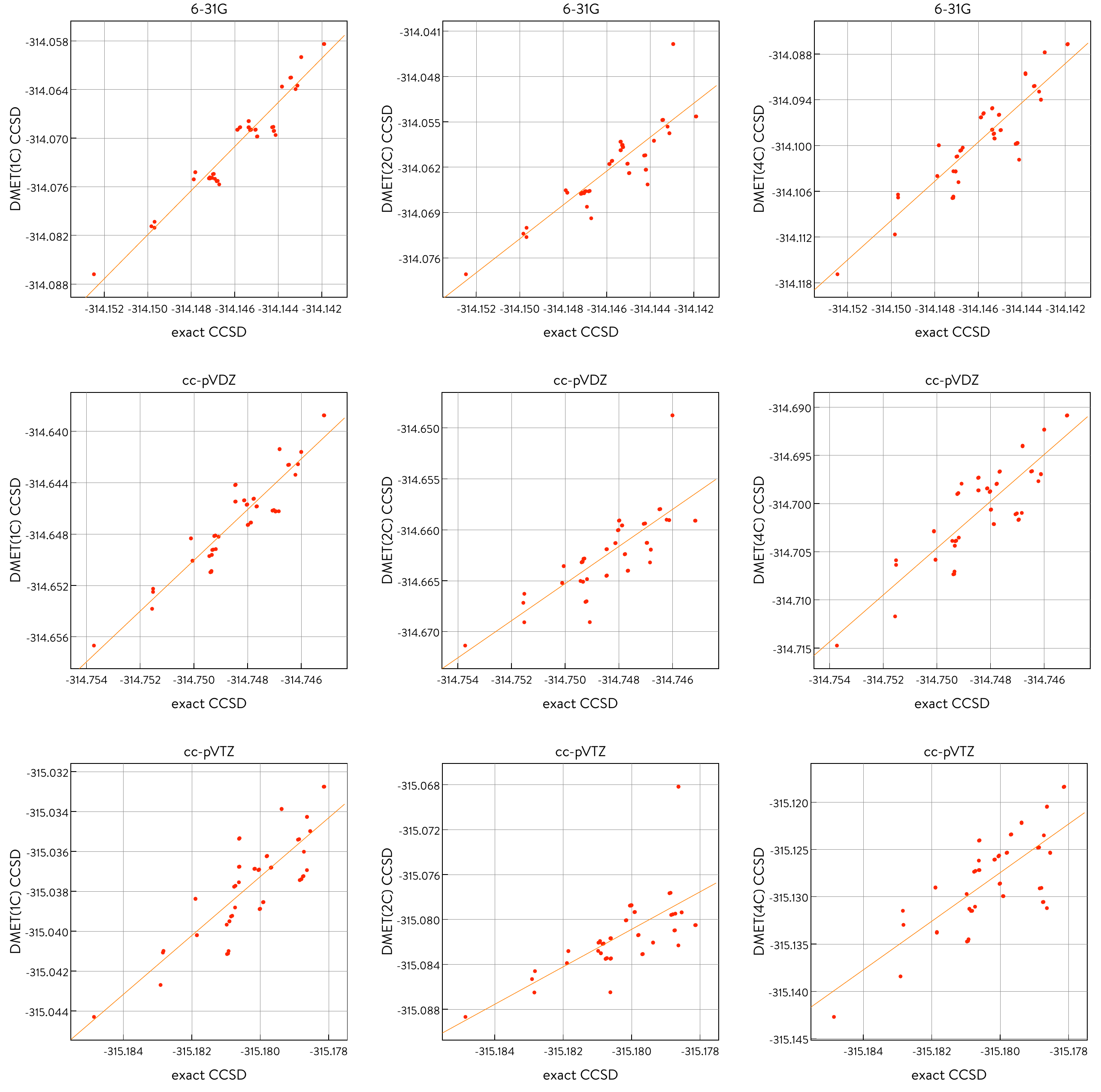}
\caption{Scatter plots of CCSD energy with DMET and the exact CCSD energy. ``1C'', ``2C'', and ``4C'' in  parentheses indicate that the DMET calculation was performed with fragments involving one-, two-, and four-carbon atoms (and the hydrogen atoms bonded to them), respectively. The fragment size increases along the horizontal axis from left to right. The basis set size increases along the vertical axis from top to bottom.}
\label{fig:DMET_9comb}
\end{figure}

In the previous assessment, we used a small fragment size that involves only one carbon atom (and the hydrogen atoms bonded to it) and the 6-31G basis set. In this subsection, we vary the fragment size as well as the basis set size in order to identify the most efficient combination between the fragment size and the basis set size for octane. For the fragment size, we consider the fragments that involve one-, two-, and four-carbon atoms (and the hydrogen atoms bonded to them), and for the basis set, we consider 6-31G, cc-pVDZ, and cc-pVTZ. Therefore, we have nine combinations in total. The results are summarized in Fig.~\ref{fig:DMET_9comb} and in Table~\ref{tab:DMET_9comb}. Overall, DMET performs reasonably well for all of the combinations, and we observe a general trend where the increase in the fragment size increases the accuracy in terms of MAD. However, there seems to be a trend that both the Pearson correlation coefficient and Spearman's rank correlation coefficient drop slightly as the fragment size increases. In addition, it seems that using large basis sets does not always yield better performance in the present case. As a result, the combination of the smallest fragment size (1C) and the smallest basis set size (6-31G) seems to be the most efficient combination. In order to make it clearer, we defined the efficiency index $I^{\text{eff}}$ as $(\rho_{\text{P}} \times \rho_{\text{S}})/\text{MAD}/\text{Ratio}$, where larger $I^{\text{eff}}$ means more efficiency. Then, it was confirmed that the combination of
1C and 6-31G has the highest $I^{\text{eff}}$ among the nine combinations. In the following subsection, we examine DMET with this combination to see if it works for even larger molecules.


\begin{table}[h]
\caption {Ratio, MAD, $\rho_{\text{P}}$, $\rho_{\text{S}}$, and $I^{\text{eff}}$ of DMET calculations for the nine combinations of three fragment sizes and three basis sets for octane.} \label{tab:DMET_9comb}
\setlength{\tabcolsep}{1em}
\begin{tabular}{ c | c | c | c }
\hline
   & 1C & 2C & 4C \\
\hline
  6-31G & \begin{small}\begin{tabular}{@{}c@{}}Ratio = 46/216 \\ MAD = 0.075 \\ $\rho_{\text{P}}$ = 0.96 \\ $\rho_{\text{S}}$ = 0.86 \\ $I^{\text{eff}}$ = \textbf{52} \end{tabular}\end{small} &
                \begin{small}\begin{tabular}{@{}c@{}}Ratio = 86/216 \\ MAD = 0.084 \\ $\rho_{\text{P}}$ = 0.87 \\ $\rho_{\text{S}}$ = 0.87 \\ $I^{\text{eff}}$ = \textbf{23} \end{tabular}\end{small}  &
                \begin{small}\begin{tabular}{@{}c@{}}Ratio = 158/216 \\ MAD = 0.047 \\ $\rho_{\text{P}}$ = 0.89 \\ $\rho_{\text{S}}$ = 0.84 \\ $I^{\text{eff}}$ = \textbf{22} \end{tabular}\end{small}   \\
 \hline
  cc-pVDZ & \begin{small}\begin{tabular}{@{}c@{}}Ratio = 74/404 \\ MAD = 0.10 \\ $\rho_{\text{P}}$ = 0.93 \\ $\rho_{\text{S}}$ = 0.88 \\ $I^{\text{eff}}$ = \textbf{45} \end{tabular}\end{small} &
                \begin{small}\begin{tabular}{@{}c@{}}Ratio = 136/404 \\ MAD = 0.086 \\ $\rho_{\text{P}}$ = 0.77 \\ $\rho_{\text{S}}$ = 0.81 \\ $I^{\text{eff}}$ = \textbf{22} \end{tabular}\end{small}  &
                \begin{small}\begin{tabular}{@{}c@{}}Ratio = 252/404 \\ MAD = 0.048 \\ $\rho_{\text{P}}$ = 0.85 \\ $\rho_{\text{S}}$ = 0.83 \\ $I^{\text{eff}}$ = \textbf{24} \end{tabular}\end{small}   \\
 \hline
  cc-pVTZ & \begin{small}\begin{tabular}{@{}c@{}}Ratio = 160/984 \\ MAD = 0.14 \\ $\rho_{\text{P}}$ = 0.83 \\ $\rho_{\text{S}}$ = 0.81 \\ $I^{\text{eff}}$ = \textbf{30} \end{tabular}\end{small} &
                \begin{small}\begin{tabular}{@{}c@{}}Ratio = 290/984 \\ MAD = 0.099 \\ $\rho_{\text{P}}$ = 0.66 \\ $\rho_{\text{S}}$ = 0.70 \\ $I^{\text{eff}}$ = \textbf{16} \end{tabular}\end{small}  &
                \begin{small}\begin{tabular}{@{}c@{}}Ratio = 542/984 \\ MAD = 0.052 \\ $\rho_{\text{P}}$ = 0.74 \\ $\rho_{\text{S}}$ = 0.70 \\ $I^{\text{eff}}$ = \textbf{18} \end{tabular}\end{small}   \\
  \hline
\end{tabular}
\end{table}


\subsection{Performance of the density matrix embedding theory for decane and dodecane molecules}
\begin{figure}
\includegraphics[width=9cm]{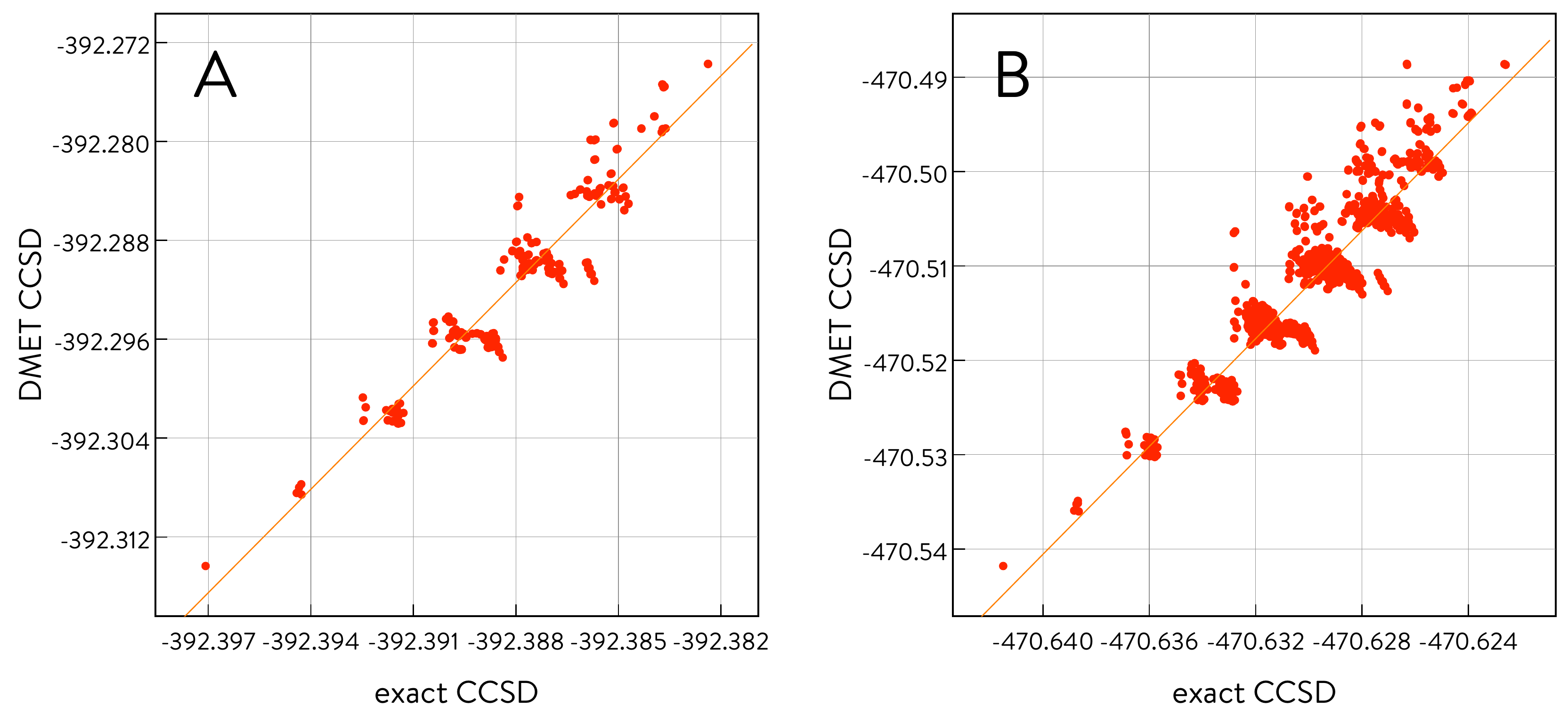}
\caption{Scatter plots of CCSD energy with DMET and the exact CCSD energy for decane (A) and dodecane (B).}
\label{fig:decane_dodecane}
\end{figure}

In this subsection, we consider decane and dodecane molecules and examine the performance of DMET with the combination of fragments involving one carbon atom (and the hydrogen atoms bonded to it) and the 6-31G basis set that provided the highest efficiency index in the previous section. We can see from Fig.~\ref{fig:decane_dodecane} and Table~\ref{tab:decane_dodecane} that DMET works reasonably well even when  the molecular size and the number of conformers were increased (294 and 1694 conformers, respectively, for decane and dodecane), indicating that DMET has strong potential to help realize practical quantum chemistry simulations on near-term quantum devices.

\begin{table}
\caption {Ratio, MAD, $\rho_{\text{P}}$, and $\rho_{\text{S}}$ for decane and dodecane.} \label{tab:decane_dodecane}
\setlength{\tabcolsep}{1em}
\begin{tabular}{ c | c c }
\hline
   & Decane & Dodecane \\
\hline
  Ratio & 46/268 & 46/320  \\
  MAD & 0.097 & 0.12  \\
  $\rho_{P}$ & 0.94 & 0.92   \\
  $\rho_{S}$ & 0.88 & 0.89    \\
  \hline
\end{tabular}
\end{table}

\section{Sampling Error Estimation}
\label{sec:sampling}

In quantum chemistry simulations on quantum devices, we need to take into account sampling errors: the expectation values of observables such as the Hamiltonian and the electron number operator will be distributed around the desired values with finite variances, which become zero only in the large sampling number limit.
In this section, we investigate how these sampling errors affect our results.

\subsection{Sampling error in the fragment molecular-orbital and divide-and-conquer methods}
\label{subsec:Error_FMODC}

In the FMO and DC methods, the correlation energy of each fragment is computed independently from the rest of the fragments.
To explain, let us consider the energy estimation in the FMO method.
For each fragment energy estimation of $E_{I}$ or $E_{IJ}$, there is a sampling error.
Let us denote the fragment Hamiltonian $\hat{H}_{I}$ as
\bea
\hat{H}_{I}=\sum_{i,j,\ldots} h^{I}_{i,j,\ldots}\sigma_{i}\otimes \sigma_{j}\otimes \cdots\,,
\eea
with Pauli matrices $\sigma_i$. Since the expectation value of the tensor product of Pauli matrices ($\sigma_{i}\otimes \sigma_{j}\otimes \cdots$) is in $[-1,1]$,
the variance of the expectation value of each term $h^{I}_{i,j,\ldots}\sigma_{i}\otimes \sigma_{j}\otimes$ is bounded by $|h^{I}_{i,j,\ldots}|^2$.
Therefore, the variance of the expectation value of $\hat{H}_{I}$ is bounded as
\bea
\text{Var}(\langle \hat{H}_{I} \rangle) \le \sum_{i,j,\ldots} {|h^{I}_{i,j,\ldots}|^2\over M_{I}}.
\eea
Here, $M_I$ is the number of samples used to estimate $\langle \hat{H}_{I} \rangle$.
We assume that estimation of each term is done independently so that the covariance is zero.
If we group the tensor product of Pauli matrices so that all the elements of each group commute each other, then we can estimate the expectation values of all the terms in each group at the same time. In this case, the covariance will take non-zero values.
We repeat a similar analysis for each fragment Hamiltonian $H^{I}$ and $H^{IJ}$.
The variance of the total energy is then
\bea
\text{Var}(\langle \hat{H}^{\text{FMO}}\rangle) \le \sum_{A\in \{I, IJ\}}\sum_{i,j,\ldots} {|h^{A}_{i,j,\ldots}|^2\over M_{A}}.
\eea
If we fix the number of samples for all of the fragments to $M$, the variance is inversely proportional to $M$.

\subsection{Sampling error in the density matrix embedding theory}
\label{subsec:Error_DMET}

In the quantum simulation of DMET, we estimate the fragment energy $\langle \hat{H}_I \rangle$ as well as the electron number $\langle \hat{N}_I \rangle = \sum_{r\in A_{I}} \langle \hat{a}^{\dagger}_{r}\hat{a}_{r} \rangle$ by using the VQE algorithm. Note that in most cases, the terms $\hat{a}^{\dagger}_{r}\hat{a}_{r}$ are included in $\hat{H}_I$. In those cases, there is no need to measure $\hat{N}_I$ independently. 

Now the sampling error comes in the measurements of the fragment Hamiltonian as well as the number of electrons for each fragment.
The variance for the expectation value of the electron number in the fragment $I$ is bounded as
\bea
\text{Var}(\langle \hat{N}_{I} \rangle) \le  {N^{\text{so}}_{I}\over M_{I}},
\eea
where $N^{\text{so}}_{I}$ is the number of spin orbitals in the fragment $I$.
We change the chemical potential based on the measured value of $N=\sum_{I}\langle \hat{N}_{I} \rangle$.
Due to the sampling error, the fragment Hamiltonian we use contains an error in its form.
This is why the sampling errors in DMET are non-trivially involved in the resulting total energy and, in general, they would no longer have a simple form proportional to the inverse of $M$.

In the present error estimation, we hypothesize the situation in which we run DMET starting from the optimal chemical potential, so that the error propagation between fragments does not occur. In this hypothetical situation, the sampling error analysis becomes exactly the same as in the case of the FMO and DC methods. The sampling error analysis of DMET will need to be revisited in the future, taking into account the error propagation.

\begin{figure}
\includegraphics[width=.75\textwidth]{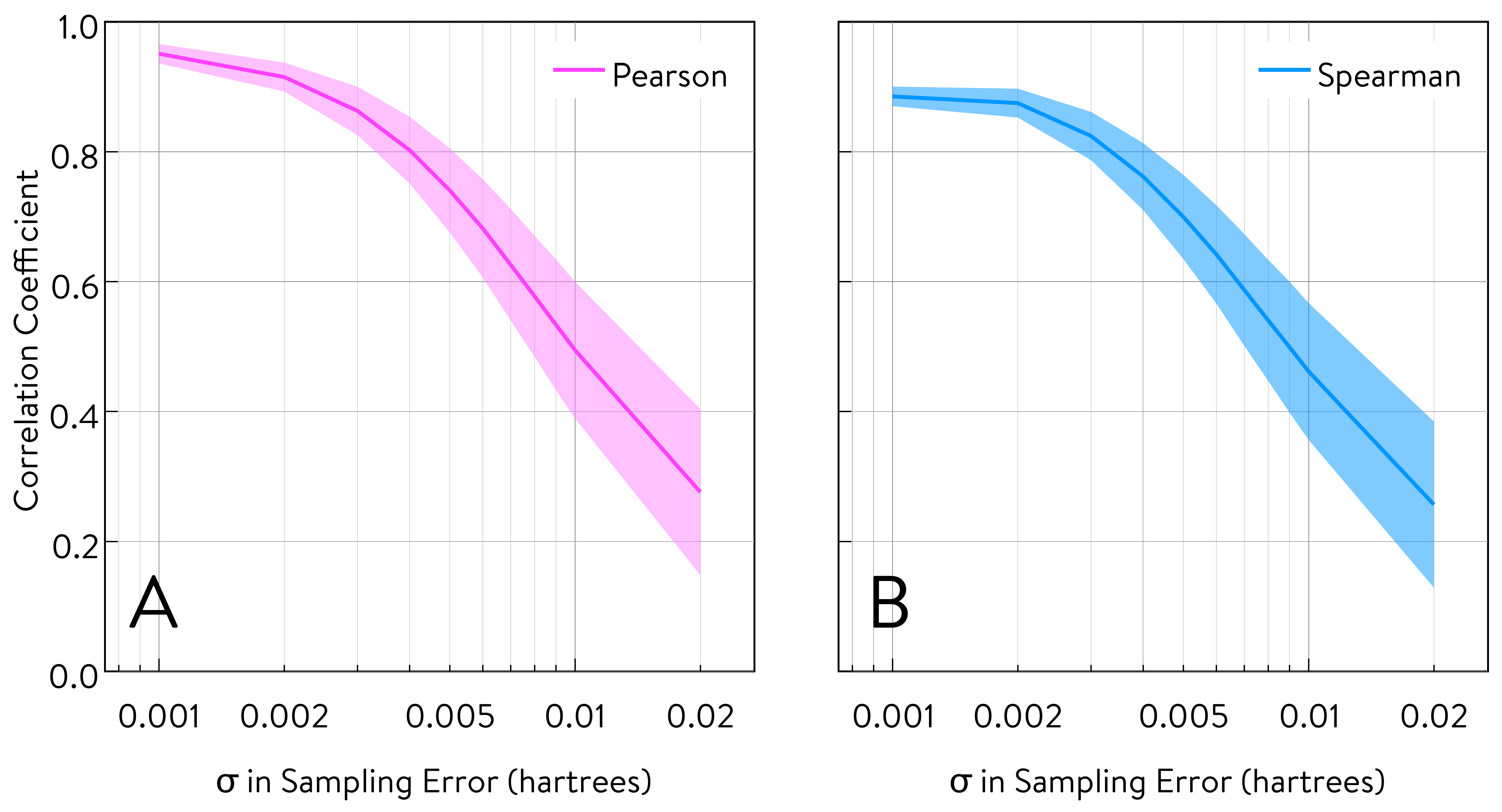}
\caption{Pearson and Spearman's rank correlation coefficients (A and B, respectively) for different sampling error levels for octane. The results are based on the bootstrap approach, where a normal distribution is assumed and its mean is equal to the ideal, error-free case. The standard deviation $\sigma$ for the sampling error varies from 0.001 to 0.02 hartrees. The solid line shows the mean of the correlation coefficients, and the shaded region indicates the standard deviation of the correlation coefficients from the mean.}
\label{fig:corrVSnoise}
\end{figure}

\subsection{Error tolerance}
\label{sec:errorTol}
Based on the above analysis, we examine how the correlation coefficients $\rho_{\text{P}}$ and $\rho_{\text{S}}$ change depending on the sampling error represented by the standard deviation values.
For the error estimation, we consider the DMET result for octane (Fig.~\ref{fig:PD_comparison}C) as an example of the PD result and as the ideal case in which there is no sampling error.
We assume that the PD energies measured many times with the sampling error will form a normal distribution with their mean equal to the ideal, error-free case (this assumption holds if the sampling error does not propagate between the fragments). In the present analysis, we consider the range of standard deviation $\sigma$ between $0.001$ to $0.02$ hartrees.
Around each point in Fig.~\ref{fig:PD_comparison}C, we draw 20,000 samples that follow the normal distribution with a target standard deviation $\sigma$, and then calculate the mean correlation value, and the standard deviation of correlation values. We note that, in the interest of simplicity, the same $\sigma$ was assumed for all points in Fig.~\ref{fig:PD_comparison}C. As can be seen in Fig.~\ref{fig:corrVSnoise}, the correlation between the PD energy and the exact energy decreases gradually as the sampling error level increases. For example, $\rho_{\text{P}}$ drops from 0.96 to 0.74 and $\rho_{\text{S}}$ drops from 0.86 to 0.70 when $\sigma$ becomes 0.005 hartrees. Therefore, it is essential to take into consideration the sampling error when we implement PD techniques on near-term quantum devices and utilize them for practical applications.

\section{Concluding remarks}
\label{sec: summary}

Quantum chemistry simulation on a quantum device hold a significant potential to accelerate the materials design process by accurately predicting the various functionalities of a given material. However, the power of quantum computing depends on various factors of the device, such as the gate depth and the number of qubits, and, although recent progress in the development of quantum devices has been remarkable, it is unlikely that we can directly perform quantum chemistry simulations of the size of molecules relevant to industry on near-term quantum hardware.

In the present study, we have proposed an alternative strategy for efficiently simulating large molecules on quantum devices based on the hybrid quantum--classical framework of leveraging problem decomposition (PD) techniques in quantum chemistry. PD techniques are a set of computational technique for decomposing a target molecular system into smaller subsystems requiring fewer computational resources. Using these techniques, we estimate the electronic structure of each subsystem, and combine the subsytems to obtain the electronic structure of the whole system.

We investigated three popular PD techniques for the hybrid framework: the fragment molecular-orbital (FMO) method, the divide-and-conquer (DC) approach, and the density matrix embedding theory (DMET). To examine these PD techniques, we targeted the straight-chain octane molecule, generated the 52 conformations octane can take, and assessed the performance of PD techniques from the viewpoints of: 1) the ratio between the number of qubits required for PD and that for the full system; 2) the mean absolute deviation; and 3) the Pearson correlation coefficient and Spearman's rank correlation coefficient. All of the PD techniques considered here showed good performance with the different range of computational resource requirements. Our assessment suggested that the FMO and DC methods would become interesting candidates to implement in our hybrid framework if we were given access to quantum hardware equipped with more than 100 qubits. The analysis also demonstrated that DMET would show the efficiency on quantum hardware with 50--100 qubits, and we observed reasonable performance of DMET when the molecular size and the number of conformers is greater as in the case of decane (294 conformers) and dodecane (1694 conformers).
 
However, the assessments above were all based on classical quantum chemistry calculations performed on classical computers, and in the execution of quantum chemistry simulations on a quantum device, the measurement of expectation values can inherently involve sampling error. Therefore, we next investigated how the sampling error changes the predictive performance of PD techniques based on a simple error analysis, and observed that the performance of the PD technique gradually decreases as the sampling-error level increases. The present analysis addresses the importance of taking into consideration the sampling error when developing a hybrid framework with PD techniques, and the necessity of more-comprehensive error analysis for PD techniques in future work.

One of our main focuses was on the reduction of the number of qubits when decomposing the full molecular system into smaller subsystems. However, we note that the quantum gate depth, the other important metric of quantum hardware performance, is, in general, proportional to the size of molecule we simulate. Therefore, we expect that the present framework can also help to make the quantum circuit shallower. In addition, our hybrid framework based on PD techniques is able to be integrated into any other ideas for reducing circuit depth. In this way, we envision that our framework contributes to the paving of a road towards practical quantum chemistry simulation at a scale close to the size of molecules relevant to industry on near-term quantum hardware.

\section*{Acknowledgements}
The authors would like to thank Marko Bucyk for editing the manuscript and Garnet Kin-Lic Chan for reviewing the paper and for valuable feedback. We are also grateful to Al\'{a}n Aspuru-Guzik and Kausar N. Samli for useful discussions. This work was supported by 1QBit.

\bibliographystyle{apsrev4-1}
\bibliography{PD_arXiv}

\begin{thebibliography}{47}%
\makeatletter
\providecommand \@ifxundefined [1]{%
 \@ifx{#1\undefined}
}%
\providecommand \@ifnum [1]{%
 \ifnum #1\expandafter \@firstoftwo
 \else \expandafter \@secondoftwo
 \fi
}%
\providecommand \@ifx [1]{%
 \ifx #1\expandafter \@firstoftwo
 \else \expandafter \@secondoftwo
 \fi
}%
\providecommand \natexlab [1]{#1}%
\providecommand \enquote  [1]{``#1''}%
\providecommand \bibnamefont  [1]{#1}%
\providecommand \bibfnamefont [1]{#1}%
\providecommand \citenamefont [1]{#1}%
\providecommand \href@noop [0]{\@secondoftwo}%
\providecommand \href [0]{\begingroup \@sanitize@url \@href}%
\providecommand \@href[1]{\@@startlink{#1}\@@href}%
\providecommand \@@href[1]{\endgroup#1\@@endlink}%
\providecommand \@sanitize@url [0]{\catcode `\\12\catcode `\$12\catcode
  `\&12\catcode `\#12\catcode `\^12\catcode `\_12\catcode `\%12\relax}%
\providecommand \@@startlink[1]{}%
\providecommand \@@endlink[0]{}%
\providecommand \url  [0]{\begingroup\@sanitize@url \@url }%
\providecommand \@url [1]{\endgroup\@href {#1}{\urlprefix }}%
\providecommand \urlprefix  [0]{URL }%
\providecommand \Eprint [0]{\href }%
\providecommand \doibase [0]{http://dx.doi.org/}%
\providecommand \selectlanguage [0]{\@gobble}%
\providecommand \bibinfo  [0]{\@secondoftwo}%
\providecommand \bibfield  [0]{\@secondoftwo}%
\providecommand \translation [1]{[#1]}%
\providecommand \BibitemOpen [0]{}%
\providecommand \bibitemStop [0]{}%
\providecommand \bibitemNoStop [0]{.\EOS\space}%
\providecommand \EOS [0]{\spacefactor3000\relax}%
\providecommand \BibitemShut  [1]{\csname bibitem#1\endcsname}%
\let\auto@bib@innerbib\@empty
\bibitem [{\citenamefont {Head-Gordon}\ and\ \citenamefont
  {Artacho}(2008)}]{Head-Gordon-2008-58}%
  \BibitemOpen
  \bibfield  {author} {\bibinfo {author} {\bibfnamefont {M.}~\bibnamefont
  {Head-Gordon}}\ and\ \bibinfo {author} {\bibfnamefont {E.}~\bibnamefont
  {Artacho}},\ }\href {\doibase 10.1063/1.2911179} {\bibfield  {journal}
  {\bibinfo  {journal} {Physics Today}\ }\textbf {\bibinfo {volume} {61}},\
  \bibinfo {pages} {58} (\bibinfo {year} {2008})}\BibitemShut {NoStop}%
\bibitem [{\citenamefont {Feynman}(1982)}]{Feynman:1982}%
  \BibitemOpen
  \bibfield  {author} {\bibinfo {author} {\bibfnamefont {R.}~\bibnamefont
  {Feynman}},\ }\href@noop {} {\bibfield  {journal} {\bibinfo  {journal}
  {International Journal of Theoretical Physics}\ }\textbf {\bibinfo {volume}
  {21}},\ \bibinfo {pages} {467} (\bibinfo {year} {1982})}\BibitemShut
  {NoStop}%
\bibitem [{\citenamefont {Abrams}\ and\ \citenamefont
  {Lloyd}(1997)}]{AbramsLloyd:PEA1997}%
  \BibitemOpen
  \bibfield  {author} {\bibinfo {author} {\bibfnamefont {D.~S.}\ \bibnamefont
  {Abrams}}\ and\ \bibinfo {author} {\bibfnamefont {S.}~\bibnamefont {Lloyd}},\
  }\href {\doibase 10.1103/PhysRevLett.79.2586} {\bibfield  {journal} {\bibinfo
   {journal} {Phys. Rev. Lett.}\ }\textbf {\bibinfo {volume} {79}},\ \bibinfo
  {pages} {2586} (\bibinfo {year} {1997})}\BibitemShut {NoStop}%
\bibitem [{\citenamefont {Aspuru-Guzik}\ \emph {et~al.}(2005)\citenamefont
  {Aspuru-Guzik}, \citenamefont {Dutoi}, \citenamefont {Love},\ and\
  \citenamefont {Head-Gordon}}]{Aspuru-Guzik:2005aa}%
  \BibitemOpen
  \bibfield  {author} {\bibinfo {author} {\bibfnamefont {A.}~\bibnamefont
  {Aspuru-Guzik}}, \bibinfo {author} {\bibfnamefont {A.~D.}\ \bibnamefont
  {Dutoi}}, \bibinfo {author} {\bibfnamefont {P.~J.}\ \bibnamefont {Love}}, \
  and\ \bibinfo {author} {\bibfnamefont {M.}~\bibnamefont {Head-Gordon}},\
  }\href {http://science.sciencemag.org/content/309/5741/ 1704.abstract}
  {\bibfield  {journal} {\bibinfo  {journal} {Science}\ }\textbf {\bibinfo
  {volume} {309}},\ \bibinfo {pages} {1704} (\bibinfo {year}
  {2005})}\BibitemShut {NoStop}%
\bibitem [{\citenamefont {Lanyon}\ \emph {et~al.}(2009)\citenamefont {Lanyon},
  \citenamefont {Whitfield}, \citenamefont {Gillet}, \citenamefont {Goggin},
  \citenamefont {Almeida}, \citenamefont {Kassal}, \citenamefont {Biamonte},
  \citenamefont {Mohseni}, \citenamefont {Powell}, \citenamefont {Barbieri},
  \citenamefont {Aspuru-Guzik},\ and\ \citenamefont {White}}]{LanyonQCex}%
  \BibitemOpen
  \bibfield  {author} {\bibinfo {author} {\bibfnamefont {B.~P.}\ \bibnamefont
  {Lanyon}}, \bibinfo {author} {\bibfnamefont {J.~D.}\ \bibnamefont
  {Whitfield}}, \bibinfo {author} {\bibfnamefont {G.~G.}\ \bibnamefont
  {Gillet}}, \bibinfo {author} {\bibfnamefont {M.~E.}\ \bibnamefont {Goggin}},
  \bibinfo {author} {\bibfnamefont {M.~P.}\ \bibnamefont {Almeida}}, \bibinfo
  {author} {\bibfnamefont {I.}~\bibnamefont {Kassal}}, \bibinfo {author}
  {\bibfnamefont {J.~D.}\ \bibnamefont {Biamonte}}, \bibinfo {author}
  {\bibfnamefont {M.}~\bibnamefont {Mohseni}}, \bibinfo {author} {\bibfnamefont
  {B.~J.}\ \bibnamefont {Powell}}, \bibinfo {author} {\bibfnamefont
  {M.}~\bibnamefont {Barbieri}}, \bibinfo {author} {\bibfnamefont
  {A.}~\bibnamefont {Aspuru-Guzik}}, \ and\ \bibinfo {author} {\bibfnamefont
  {A.~G.}\ \bibnamefont {White}},\ }\href {\doibase 10.1038/nchem.483} {\
  (\bibinfo {year} {2009}),\ 10.1038/nchem.483},\ \Eprint
  {http://arxiv.org/abs/arXiv:0905.0887} {arXiv:0905.0887} \BibitemShut
  {NoStop}%
\bibitem [{\citenamefont {Du}\ \emph {et~al.}(2010)\citenamefont {Du},
  \citenamefont {Xu}, \citenamefont {Peng}, \citenamefont {Wang}, \citenamefont
  {Wu},\ and\ \citenamefont {Lu}}]{PhysRevLett.104.030502}%
  \BibitemOpen
  \bibfield  {author} {\bibinfo {author} {\bibfnamefont {J.}~\bibnamefont
  {Du}}, \bibinfo {author} {\bibfnamefont {N.}~\bibnamefont {Xu}}, \bibinfo
  {author} {\bibfnamefont {X.}~\bibnamefont {Peng}}, \bibinfo {author}
  {\bibfnamefont {P.}~\bibnamefont {Wang}}, \bibinfo {author} {\bibfnamefont
  {S.}~\bibnamefont {Wu}}, \ and\ \bibinfo {author} {\bibfnamefont
  {D.}~\bibnamefont {Lu}},\ }\href {\doibase 10.1103/PhysRevLett.104.030502}
  {\bibfield  {journal} {\bibinfo  {journal} {Phys. Rev. Lett.}\ }\textbf
  {\bibinfo {volume} {104}},\ \bibinfo {pages} {030502} (\bibinfo {year}
  {2010})}\BibitemShut {NoStop}%
\bibitem [{\citenamefont {Wang}\ \emph {et~al.}(2014)\citenamefont {Wang},
  \citenamefont {Dolde}, \citenamefont {Biamonte}, \citenamefont {Babbush},
  \citenamefont {Bergholm}, \citenamefont {Yang}, \citenamefont {Jakobi},
  \citenamefont {Neumann}, \citenamefont {Aspuru-Guzik}, \citenamefont
  {Whitfield},\ and\ \citenamefont {Wrachtrup}}]{Wang2014}%
  \BibitemOpen
  \bibfield  {author} {\bibinfo {author} {\bibfnamefont {Y.}~\bibnamefont
  {Wang}}, \bibinfo {author} {\bibfnamefont {F.}~\bibnamefont {Dolde}},
  \bibinfo {author} {\bibfnamefont {J.}~\bibnamefont {Biamonte}}, \bibinfo
  {author} {\bibfnamefont {R.}~\bibnamefont {Babbush}}, \bibinfo {author}
  {\bibfnamefont {V.}~\bibnamefont {Bergholm}}, \bibinfo {author}
  {\bibfnamefont {S.}~\bibnamefont {Yang}}, \bibinfo {author} {\bibfnamefont
  {I.}~\bibnamefont {Jakobi}}, \bibinfo {author} {\bibfnamefont
  {P.}~\bibnamefont {Neumann}}, \bibinfo {author} {\bibfnamefont
  {A.}~\bibnamefont {Aspuru-Guzik}}, \bibinfo {author} {\bibfnamefont {J.~D.}\
  \bibnamefont {Whitfield}}, \ and\ \bibinfo {author} {\bibfnamefont
  {J.}~\bibnamefont {Wrachtrup}},\ }\href {\doibase 10.1021/acsnano.5b01651} {\
   (\bibinfo {year} {2014}),\ 10.1021/acsnano.5b01651},\ \Eprint
  {http://arxiv.org/abs/arXiv:1405.2696} {arXiv:1405.2696} \BibitemShut
  {NoStop}%
\bibitem [{\citenamefont {Paesani}\ \emph {et~al.}(2017)\citenamefont
  {Paesani}, \citenamefont {Gentile}, \citenamefont {Santagati}, \citenamefont
  {Wang}, \citenamefont {Wiebe}, \citenamefont {Tew}, \citenamefont {O'Brien},\
  and\ \citenamefont {Thompson}}]{PhysRevLett.118.100503}%
  \BibitemOpen
  \bibfield  {author} {\bibinfo {author} {\bibfnamefont {S.}~\bibnamefont
  {Paesani}}, \bibinfo {author} {\bibfnamefont {A.~A.}\ \bibnamefont
  {Gentile}}, \bibinfo {author} {\bibfnamefont {R.}~\bibnamefont {Santagati}},
  \bibinfo {author} {\bibfnamefont {J.}~\bibnamefont {Wang}}, \bibinfo {author}
  {\bibfnamefont {N.}~\bibnamefont {Wiebe}}, \bibinfo {author} {\bibfnamefont
  {D.~P.}\ \bibnamefont {Tew}}, \bibinfo {author} {\bibfnamefont {J.~L.}\
  \bibnamefont {O'Brien}}, \ and\ \bibinfo {author} {\bibfnamefont {M.~G.}\
  \bibnamefont {Thompson}},\ }\href {\doibase 10.1103/PhysRevLett.118.100503}
  {\bibfield  {journal} {\bibinfo  {journal} {Phys. Rev. Lett.}\ }\textbf
  {\bibinfo {volume} {118}},\ \bibinfo {pages} {100503} (\bibinfo {year}
  {2017})}\BibitemShut {NoStop}%
\bibitem [{\citenamefont {Peruzzo}\ \emph {et~al.}(2014)\citenamefont
  {Peruzzo}, \citenamefont {McClean}, \citenamefont {Shadbolt}, \citenamefont
  {Yung}, \citenamefont {Zhou}, \citenamefont {Love}, \citenamefont
  {Aspuru-Guzik},\ and\ \citenamefont {O'Brien}}]{Peruzzo:2014aa}%
  \BibitemOpen
  \bibfield  {author} {\bibinfo {author} {\bibfnamefont {A.}~\bibnamefont
  {Peruzzo}}, \bibinfo {author} {\bibfnamefont {J.}~\bibnamefont {McClean}},
  \bibinfo {author} {\bibfnamefont {P.}~\bibnamefont {Shadbolt}}, \bibinfo
  {author} {\bibfnamefont {M.-H.}\ \bibnamefont {Yung}}, \bibinfo {author}
  {\bibfnamefont {X.-Q.}\ \bibnamefont {Zhou}}, \bibinfo {author}
  {\bibfnamefont {P.~J.}\ \bibnamefont {Love}}, \bibinfo {author}
  {\bibfnamefont {A.}~\bibnamefont {Aspuru-Guzik}}, \ and\ \bibinfo {author}
  {\bibfnamefont {J.~L.}\ \bibnamefont {O'Brien}},\ }\href
  {http://dx.doi.org/10.1038/ncomms5213} {\bibfield  {journal} {\bibinfo
  {journal} {Nat. Commun.}\ }\textbf {\bibinfo {volume} {5}},\ \bibinfo {pages}
  {4213 EP } (\bibinfo {year} {2014})}\BibitemShut {NoStop}%
\bibitem [{\citenamefont {Yung}\ \emph {et~al.}(2014)\citenamefont {Yung},
  \citenamefont {Casanova}, \citenamefont {Mezzacapo}, \citenamefont {McClean},
  \citenamefont {Lamata}, \citenamefont {Aspuru-Guzik},\ and\ \citenamefont
  {Solano}}]{VQE2}%
  \BibitemOpen
  \bibfield  {author} {\bibinfo {author} {\bibfnamefont {M.~H.}\ \bibnamefont
  {Yung}}, \bibinfo {author} {\bibfnamefont {J.}~\bibnamefont {Casanova}},
  \bibinfo {author} {\bibfnamefont {A.}~\bibnamefont {Mezzacapo}}, \bibinfo
  {author} {\bibfnamefont {J.}~\bibnamefont {McClean}}, \bibinfo {author}
  {\bibfnamefont {L.}~\bibnamefont {Lamata}}, \bibinfo {author} {\bibfnamefont
  {A.}~\bibnamefont {Aspuru-Guzik}}, \ and\ \bibinfo {author} {\bibfnamefont
  {E.}~\bibnamefont {Solano}},\ }\href {http://dx.doi.org/10.1038/srep03589}
  {\bibfield  {journal} {\bibinfo  {journal} {Scientific Reports}\ }\textbf
  {\bibinfo {volume} {4}},\ \bibinfo {pages} {3589 EP } (\bibinfo {year}
  {2014})}\BibitemShut {NoStop}%
\bibitem [{\citenamefont {McClean}\ \emph {et~al.}(2016)\citenamefont
  {McClean}, \citenamefont {Romero}, \citenamefont {Babbush},\ and\
  \citenamefont {Aspuru-Guzik}}]{VQE3}%
  \BibitemOpen
  \bibfield  {author} {\bibinfo {author} {\bibfnamefont {J.~R.}\ \bibnamefont
  {McClean}}, \bibinfo {author} {\bibfnamefont {J.}~\bibnamefont {Romero}},
  \bibinfo {author} {\bibfnamefont {R.}~\bibnamefont {Babbush}}, \ and\
  \bibinfo {author} {\bibfnamefont {A.}~\bibnamefont {Aspuru-Guzik}},\ }\href
  {http://stacks.iop.org/1367-2630/18/i=2/a=023023} {\bibfield  {journal}
  {\bibinfo  {journal} {New Journal of Physics}\ }\textbf {\bibinfo {volume}
  {18}},\ \bibinfo {pages} {023023} (\bibinfo {year} {2016})}\BibitemShut
  {NoStop}%
\bibitem [{\citenamefont {Wecker}\ \emph
  {et~al.}(2015{\natexlab{a}})\citenamefont {Wecker}, \citenamefont {Hastings},
  \citenamefont {Wiebe}, \citenamefont {Clark}, \citenamefont {Nayak},\ and\
  \citenamefont {Troyer}}]{1506.05135}%
  \BibitemOpen
  \bibfield  {author} {\bibinfo {author} {\bibfnamefont {D.}~\bibnamefont
  {Wecker}}, \bibinfo {author} {\bibfnamefont {M.~B.}\ \bibnamefont
  {Hastings}}, \bibinfo {author} {\bibfnamefont {N.}~\bibnamefont {Wiebe}},
  \bibinfo {author} {\bibfnamefont {B.~K.}\ \bibnamefont {Clark}}, \bibinfo
  {author} {\bibfnamefont {C.}~\bibnamefont {Nayak}}, \ and\ \bibinfo {author}
  {\bibfnamefont {M.}~\bibnamefont {Troyer}},\ }\href {\doibase
  10.1103/PhysRevA.92.062318} {\  (\bibinfo {year} {2015}{\natexlab{a}}),\
  10.1103/PhysRevA.92.062318},\ \Eprint {http://arxiv.org/abs/arXiv:1506.05135}
  {arXiv:1506.05135} \BibitemShut {NoStop}%
\bibitem [{\citenamefont {Wecker}\ \emph
  {et~al.}(2015{\natexlab{b}})\citenamefont {Wecker}, \citenamefont
  {Hastings},\ and\ \citenamefont {Troyer}}]{PhysRevA.92.042303}%
  \BibitemOpen
  \bibfield  {author} {\bibinfo {author} {\bibfnamefont {D.}~\bibnamefont
  {Wecker}}, \bibinfo {author} {\bibfnamefont {M.~B.}\ \bibnamefont
  {Hastings}}, \ and\ \bibinfo {author} {\bibfnamefont {M.}~\bibnamefont
  {Troyer}},\ }\href {\doibase 10.1103/PhysRevA.92.042303} {\bibfield
  {journal} {\bibinfo  {journal} {Phys. Rev. A}\ }\textbf {\bibinfo {volume}
  {92}},\ \bibinfo {pages} {042303} (\bibinfo {year}
  {2015}{\natexlab{b}})}\BibitemShut {NoStop}%
\bibitem [{\citenamefont {Babbush}\ \emph
  {et~al.}(2017{\natexlab{a}})\citenamefont {Babbush}, \citenamefont {Wiebe},
  \citenamefont {McClean}, \citenamefont {McClain}, \citenamefont {Neven},\
  and\ \citenamefont {Chan}}]{1706.00023}%
  \BibitemOpen
  \bibfield  {author} {\bibinfo {author} {\bibfnamefont {R.}~\bibnamefont
  {Babbush}}, \bibinfo {author} {\bibfnamefont {N.}~\bibnamefont {Wiebe}},
  \bibinfo {author} {\bibfnamefont {J.}~\bibnamefont {McClean}}, \bibinfo
  {author} {\bibfnamefont {J.}~\bibnamefont {McClain}}, \bibinfo {author}
  {\bibfnamefont {H.}~\bibnamefont {Neven}}, \ and\ \bibinfo {author}
  {\bibfnamefont {G.~K.-L.}\ \bibnamefont {Chan}},\ }\href@noop {} {\enquote
  {\bibinfo {title} {Low depth quantum simulation of electronic structure},}\ }
  (\bibinfo {year} {2017}{\natexlab{a}}),\ \Eprint
  {http://arxiv.org/abs/arXiv:1706.00023} {arXiv:1706.00023} \BibitemShut
  {NoStop}%
\bibitem [{\citenamefont {Kivlichan}\ \emph {et~al.}(2017)\citenamefont
  {Kivlichan}, \citenamefont {McClean}, \citenamefont {Wiebe}, \citenamefont
  {Gidney}, \citenamefont {Aspuru-Guzik}, \citenamefont {Chan},\ and\
  \citenamefont {Babbush}}]{Kivlichan-2017-1711.04789}%
  \BibitemOpen
  \bibfield  {author} {\bibinfo {author} {\bibfnamefont {I.~D.}\ \bibnamefont
  {Kivlichan}}, \bibinfo {author} {\bibfnamefont {J.}~\bibnamefont {McClean}},
  \bibinfo {author} {\bibfnamefont {N.}~\bibnamefont {Wiebe}}, \bibinfo
  {author} {\bibfnamefont {C.}~\bibnamefont {Gidney}}, \bibinfo {author}
  {\bibfnamefont {A.}~\bibnamefont {Aspuru-Guzik}}, \bibinfo {author}
  {\bibfnamefont {G.~K.-L.}\ \bibnamefont {Chan}}, \ and\ \bibinfo {author}
  {\bibfnamefont {R.}~\bibnamefont {Babbush}},\ }\href@noop {} {\bibfield
  {journal} {\bibinfo  {journal} {arXiv.org}\ ,\ \bibinfo {pages}
  {arXiv:1711.04789 [quant}} (\bibinfo {year} {2017})}\BibitemShut {NoStop}%
\bibitem [{\citenamefont {Seeley}\ \emph {et~al.}(2012)\citenamefont {Seeley},
  \citenamefont {Richard},\ and\ \citenamefont {Love}}]{BK_trans_2}%
  \BibitemOpen
  \bibfield  {author} {\bibinfo {author} {\bibfnamefont {J.~T.}\ \bibnamefont
  {Seeley}}, \bibinfo {author} {\bibfnamefont {M.~J.}\ \bibnamefont {Richard}},
  \ and\ \bibinfo {author} {\bibfnamefont {P.~J.}\ \bibnamefont {Love}},\
  }\href {\doibase 10.1063/1.4768229} {\bibfield  {journal} {\bibinfo
  {journal} {The Journal of Chemical Physics}\ }\textbf {\bibinfo {volume}
  {137}},\ \bibinfo {pages} {224109} (\bibinfo {year} {2012})},\ \Eprint
  {http://arxiv.org/abs/https://doi.org/10.1063/1.4768229}
  {https://doi.org/10.1063/1.4768229} \BibitemShut {NoStop}%
\bibitem [{\citenamefont {Bravyi}\ and\ \citenamefont
  {Kitaev}(2000)}]{BK_trans_1}%
  \BibitemOpen
  \bibfield  {author} {\bibinfo {author} {\bibfnamefont {S.}~\bibnamefont
  {Bravyi}}\ and\ \bibinfo {author} {\bibfnamefont {A.}~\bibnamefont
  {Kitaev}},\ }\href {\doibase 10.1006/aphy.2002.6254} {\  (\bibinfo {year}
  {2000}),\ 10.1006/aphy.2002.6254},\ \Eprint
  {http://arxiv.org/abs/arXiv:quant-ph/0003137} {arXiv:quant-ph/0003137}
  \BibitemShut {NoStop}%
\bibitem [{\citenamefont {Bravyi}\ \emph {et~al.}(2017)\citenamefont {Bravyi},
  \citenamefont {Gambetta}, \citenamefont {Mezzacapo},\ and\ \citenamefont
  {Temme}}]{IBM_symmetry}%
  \BibitemOpen
  \bibfield  {author} {\bibinfo {author} {\bibfnamefont {S.}~\bibnamefont
  {Bravyi}}, \bibinfo {author} {\bibfnamefont {J.~M.}\ \bibnamefont
  {Gambetta}}, \bibinfo {author} {\bibfnamefont {A.}~\bibnamefont {Mezzacapo}},
  \ and\ \bibinfo {author} {\bibfnamefont {K.}~\bibnamefont {Temme}},\
  }\href@noop {} {\enquote {\bibinfo {title} {Tapering off qubits to simulate
  fermionic hamiltonians},}\ } (\bibinfo {year} {2017}),\ \Eprint
  {http://arxiv.org/abs/arXiv:1701.08213} {arXiv:1701.08213} \BibitemShut
  {NoStop}%
\bibitem [{\citenamefont {O'Malley}\ \emph {et~al.}(2016)\citenamefont
  {O'Malley}, \citenamefont {Babbush}, \citenamefont {Kivlichan}, \citenamefont
  {Romero}, \citenamefont {McClean}, \citenamefont {Barends}, \citenamefont
  {Kelly}, \citenamefont {Roushan}, \citenamefont {Tranter}, \citenamefont
  {Ding}, \citenamefont {Campbell}, \citenamefont {Chen}, \citenamefont {Chen},
  \citenamefont {Chiaro}, \citenamefont {Dunsworth}, \citenamefont {Fowler},
  \citenamefont {Jeffrey}, \citenamefont {Lucero}, \citenamefont {Megrant},
  \citenamefont {Mutus}, \citenamefont {Neeley}, \citenamefont {Neill},
  \citenamefont {Quintana}, \citenamefont {Sank}, \citenamefont {Vainsencher},
  \citenamefont {Wenner}, \citenamefont {White}, \citenamefont {Coveney},
  \citenamefont {Love}, \citenamefont {Neven}, \citenamefont {Aspuru-Guzik},\
  and\ \citenamefont {Martinis}}]{Alan-Martinis:PhysRevX.2017}%
  \BibitemOpen
  \bibfield  {author} {\bibinfo {author} {\bibfnamefont {P.~J.~J.}\
  \bibnamefont {O'Malley}}, \bibinfo {author} {\bibfnamefont {R.}~\bibnamefont
  {Babbush}}, \bibinfo {author} {\bibfnamefont {I.~D.}\ \bibnamefont
  {Kivlichan}}, \bibinfo {author} {\bibfnamefont {J.}~\bibnamefont {Romero}},
  \bibinfo {author} {\bibfnamefont {J.~R.}\ \bibnamefont {McClean}}, \bibinfo
  {author} {\bibfnamefont {R.}~\bibnamefont {Barends}}, \bibinfo {author}
  {\bibfnamefont {J.}~\bibnamefont {Kelly}}, \bibinfo {author} {\bibfnamefont
  {P.}~\bibnamefont {Roushan}}, \bibinfo {author} {\bibfnamefont
  {A.}~\bibnamefont {Tranter}}, \bibinfo {author} {\bibfnamefont
  {N.}~\bibnamefont {Ding}}, \bibinfo {author} {\bibfnamefont {B.}~\bibnamefont
  {Campbell}}, \bibinfo {author} {\bibfnamefont {Y.}~\bibnamefont {Chen}},
  \bibinfo {author} {\bibfnamefont {Z.}~\bibnamefont {Chen}}, \bibinfo {author}
  {\bibfnamefont {B.}~\bibnamefont {Chiaro}}, \bibinfo {author} {\bibfnamefont
  {A.}~\bibnamefont {Dunsworth}}, \bibinfo {author} {\bibfnamefont {A.~G.}\
  \bibnamefont {Fowler}}, \bibinfo {author} {\bibfnamefont {E.}~\bibnamefont
  {Jeffrey}}, \bibinfo {author} {\bibfnamefont {E.}~\bibnamefont {Lucero}},
  \bibinfo {author} {\bibfnamefont {A.}~\bibnamefont {Megrant}}, \bibinfo
  {author} {\bibfnamefont {J.~Y.}\ \bibnamefont {Mutus}}, \bibinfo {author}
  {\bibfnamefont {M.}~\bibnamefont {Neeley}}, \bibinfo {author} {\bibfnamefont
  {C.}~\bibnamefont {Neill}}, \bibinfo {author} {\bibfnamefont
  {C.}~\bibnamefont {Quintana}}, \bibinfo {author} {\bibfnamefont
  {D.}~\bibnamefont {Sank}}, \bibinfo {author} {\bibfnamefont {A.}~\bibnamefont
  {Vainsencher}}, \bibinfo {author} {\bibfnamefont {J.}~\bibnamefont {Wenner}},
  \bibinfo {author} {\bibfnamefont {T.~C.}\ \bibnamefont {White}}, \bibinfo
  {author} {\bibfnamefont {P.~V.}\ \bibnamefont {Coveney}}, \bibinfo {author}
  {\bibfnamefont {P.~J.}\ \bibnamefont {Love}}, \bibinfo {author}
  {\bibfnamefont {H.}~\bibnamefont {Neven}}, \bibinfo {author} {\bibfnamefont
  {A.}~\bibnamefont {Aspuru-Guzik}}, \ and\ \bibinfo {author} {\bibfnamefont
  {J.~M.}\ \bibnamefont {Martinis}},\ }\href {\doibase
  10.1103/PhysRevX.6.031007} {\bibfield  {journal} {\bibinfo  {journal} {Phys.
  Rev. X}\ }\textbf {\bibinfo {volume} {6}},\ \bibinfo {pages} {031007}
  (\bibinfo {year} {2016})}\BibitemShut {NoStop}%
\bibitem [{\citenamefont {Wishart}\ \emph {et~al.}(2018)\citenamefont
  {Wishart}, \citenamefont {Feunang}, \citenamefont {Guo}, \citenamefont {Lo},
  \citenamefont {Marcu}, \citenamefont {Grant}, \citenamefont {Sajed},
  \citenamefont {Johnson}, \citenamefont {Li}, \citenamefont {Sayeeda},
  \citenamefont {Assempour}, \citenamefont {Iynkkaran}, \citenamefont {Liu},
  \citenamefont {Maciejewski}, \citenamefont {Gale}, \citenamefont {Wilson},
  \citenamefont {Chin}, \citenamefont {Cummings}, \citenamefont {Le},
  \citenamefont {Pon}, \citenamefont {Knox},\ and\ \citenamefont
  {Wilson}}]{Wishart:2018aa}%
  \BibitemOpen
  \bibfield  {author} {\bibinfo {author} {\bibfnamefont {D.~S.}\ \bibnamefont
  {Wishart}}, \bibinfo {author} {\bibfnamefont {Y.~D.}\ \bibnamefont
  {Feunang}}, \bibinfo {author} {\bibfnamefont {A.~C.}\ \bibnamefont {Guo}},
  \bibinfo {author} {\bibfnamefont {E.~J.}\ \bibnamefont {Lo}}, \bibinfo
  {author} {\bibfnamefont {A.}~\bibnamefont {Marcu}}, \bibinfo {author}
  {\bibfnamefont {J.~R.}\ \bibnamefont {Grant}}, \bibinfo {author}
  {\bibfnamefont {T.}~\bibnamefont {Sajed}}, \bibinfo {author} {\bibfnamefont
  {D.}~\bibnamefont {Johnson}}, \bibinfo {author} {\bibfnamefont
  {C.}~\bibnamefont {Li}}, \bibinfo {author} {\bibfnamefont {Z.}~\bibnamefont
  {Sayeeda}}, \bibinfo {author} {\bibfnamefont {N.}~\bibnamefont {Assempour}},
  \bibinfo {author} {\bibfnamefont {I.}~\bibnamefont {Iynkkaran}}, \bibinfo
  {author} {\bibfnamefont {Y.}~\bibnamefont {Liu}}, \bibinfo {author}
  {\bibfnamefont {A.}~\bibnamefont {Maciejewski}}, \bibinfo {author}
  {\bibfnamefont {N.}~\bibnamefont {Gale}}, \bibinfo {author} {\bibfnamefont
  {A.}~\bibnamefont {Wilson}}, \bibinfo {author} {\bibfnamefont
  {L.}~\bibnamefont {Chin}}, \bibinfo {author} {\bibfnamefont {R.}~\bibnamefont
  {Cummings}}, \bibinfo {author} {\bibfnamefont {D.}~\bibnamefont {Le}},
  \bibinfo {author} {\bibfnamefont {A.}~\bibnamefont {Pon}}, \bibinfo {author}
  {\bibfnamefont {C.}~\bibnamefont {Knox}}, \ and\ \bibinfo {author}
  {\bibfnamefont {M.}~\bibnamefont {Wilson}},\ }\href
  {http://dx.doi.org/10.1093/nar/gkx1037} {\bibfield  {journal} {\bibinfo
  {journal} {Nucleic Acids Research}\ }\textbf {\bibinfo {volume} {46}},\
  \bibinfo {pages} {D1074} (\bibinfo {year} {2018})}\BibitemShut {NoStop}%
\bibitem [{\citenamefont {Szabo}\ and\ \citenamefont
  {Ostlund}(1996)}]{Szabo-1996}%
  \BibitemOpen
  \bibfield  {author} {\bibinfo {author} {\bibfnamefont {A.}~\bibnamefont
  {Szabo}}\ and\ \bibinfo {author} {\bibfnamefont {N.~S.}\ \bibnamefont
  {Ostlund}},\ }\href@noop {} {\emph {\bibinfo {title} {Modern Quantum
  Chemistry: Introduction to Advanced Electronic Structure Theory}}}\ (\bibinfo
   {publisher} {Dover Publications, Inc.},\ \bibinfo {year} {1996})\BibitemShut
  {NoStop}%
\bibitem [{\citenamefont {Kobayashi}\ and\ \citenamefont
  {Nakai}(2011)}]{Kobayashi:2011aa}%
  \BibitemOpen
  \bibfield  {author} {\bibinfo {author} {\bibfnamefont {M.}~\bibnamefont
  {Kobayashi}}\ and\ \bibinfo {author} {\bibfnamefont {H.}~\bibnamefont
  {Nakai}},\ }\enquote {\bibinfo {title} {Divide-and-conquer approaches to
  quantum chemistry: Theory and implementation},}\ in\ \href {\doibase
  10.1007/978-90-481-2853-2{\_}5} {\emph {\bibinfo {booktitle} {Linear-Scaling
  Techniques in Computational Chemistry and Physics: Methods and
  Applications}}},\ \bibinfo {editor} {edited by\ \bibinfo {editor}
  {\bibfnamefont {R.}~\bibnamefont {Zalesny}}, \bibinfo {editor} {\bibfnamefont
  {M.~G.}\ \bibnamefont {Papadopoulos}}, \bibinfo {editor} {\bibfnamefont
  {P.~G.}\ \bibnamefont {Mezey}}, \ and\ \bibinfo {editor} {\bibfnamefont
  {J.}~\bibnamefont {Leszczynski}}}\ (\bibinfo  {publisher} {Springer
  Netherlands},\ \bibinfo {address} {Dordrecht},\ \bibinfo {year} {2011})\ pp.\
  \bibinfo {pages} {97--127}\BibitemShut {NoStop}%
\bibitem [{\citenamefont {Imamura}\ \emph {et~al.}(1991)\citenamefont
  {Imamura}, \citenamefont {Aoki},\ and\ \citenamefont
  {Maekawa}}]{Imamura:1991aa}%
  \BibitemOpen
  \bibfield  {author} {\bibinfo {author} {\bibfnamefont {A.}~\bibnamefont
  {Imamura}}, \bibinfo {author} {\bibfnamefont {Y.}~\bibnamefont {Aoki}}, \
  and\ \bibinfo {author} {\bibfnamefont {K.}~\bibnamefont {Maekawa}},\
  }\bibfield  {booktitle} {\emph {\bibinfo {booktitle} {The Journal of Chemical
  Physics}},\ }\href {\doibase 10.1063/1.461658} {\bibfield  {journal}
  {\bibinfo  {journal} {The Journal of Chemical Physics}\ }\textbf {\bibinfo
  {volume} {95}},\ \bibinfo {pages} {5419} (\bibinfo {year}
  {1991})}\BibitemShut {NoStop}%
\bibitem [{\citenamefont {Kitaura}\ \emph {et~al.}(1999)\citenamefont
  {Kitaura}, \citenamefont {Ikeo}, \citenamefont {Asada}, \citenamefont
  {Nakano},\ and\ \citenamefont {Uebayasi}}]{Kitaura:1999aa}%
  \BibitemOpen
  \bibfield  {author} {\bibinfo {author} {\bibfnamefont {K.}~\bibnamefont
  {Kitaura}}, \bibinfo {author} {\bibfnamefont {E.}~\bibnamefont {Ikeo}},
  \bibinfo {author} {\bibfnamefont {T.}~\bibnamefont {Asada}}, \bibinfo
  {author} {\bibfnamefont {T.}~\bibnamefont {Nakano}}, \ and\ \bibinfo {author}
  {\bibfnamefont {M.}~\bibnamefont {Uebayasi}},\ }\href {\doibase
  http://dx.doi.org/10.1016/S0009-2614(99)00874-X} {\bibfield  {journal}
  {\bibinfo  {journal} {Chemical Physics Letters}\ }\textbf {\bibinfo {volume}
  {313}},\ \bibinfo {pages} {701} (\bibinfo {year} {1999})}\BibitemShut
  {NoStop}%
\bibitem [{\citenamefont {Zhang}\ and\ \citenamefont
  {Zhang}(2003)}]{Zhang:2003aa}%
  \BibitemOpen
  \bibfield  {author} {\bibinfo {author} {\bibfnamefont {D.~W.}\ \bibnamefont
  {Zhang}}\ and\ \bibinfo {author} {\bibfnamefont {J.~Z.~H.}\ \bibnamefont
  {Zhang}},\ }\bibfield  {booktitle} {\emph {\bibinfo {booktitle} {The Journal
  of Chemical Physics}},\ }\href {\doibase 10.1063/1.1591727} {\bibfield
  {journal} {\bibinfo  {journal} {The Journal of Chemical Physics}\ }\textbf
  {\bibinfo {volume} {119}},\ \bibinfo {pages} {3599} (\bibinfo {year}
  {2003})}\BibitemShut {NoStop}%
\bibitem [{\citenamefont {Gao}\ \emph {et~al.}(2010)\citenamefont {Gao},
  \citenamefont {Cembran},\ and\ \citenamefont {Mo}}]{Gao:2010aa}%
  \BibitemOpen
  \bibfield  {author} {\bibinfo {author} {\bibfnamefont {J.}~\bibnamefont
  {Gao}}, \bibinfo {author} {\bibfnamefont {A.}~\bibnamefont {Cembran}}, \ and\
  \bibinfo {author} {\bibfnamefont {Y.}~\bibnamefont {Mo}},\ }\bibfield
  {booktitle} {\emph {\bibinfo {booktitle} {Journal of Chemical Theory and
  Computation}},\ }\href {\doibase 10.1021/ct100292g} {\bibfield  {journal}
  {\bibinfo  {journal} {Journal of Chemical Theory and Computation}\ }\textbf
  {\bibinfo {volume} {6}},\ \bibinfo {pages} {2402} (\bibinfo {year}
  {2010})}\BibitemShut {NoStop}%
\bibitem [{\citenamefont {Akimov}\ and\ \citenamefont
  {Prezhdo}(2015)}]{Akimov:2015aa}%
  \BibitemOpen
  \bibfield  {author} {\bibinfo {author} {\bibfnamefont {A.~V.}\ \bibnamefont
  {Akimov}}\ and\ \bibinfo {author} {\bibfnamefont {O.~V.}\ \bibnamefont
  {Prezhdo}},\ }\bibfield  {booktitle} {\emph {\bibinfo {booktitle} {Chemical
  Reviews}},\ }\href {\doibase 10.1021/cr500524c} {\bibfield  {journal}
  {\bibinfo  {journal} {Chemical Reviews}\ }\textbf {\bibinfo {volume} {115}},\
  \bibinfo {pages} {5797} (\bibinfo {year} {2015})}\BibitemShut {NoStop}%
\bibitem [{\citenamefont {Bauer}\ \emph {et~al.}(2016)\citenamefont {Bauer},
  \citenamefont {Wecker}, \citenamefont {Millis}, \citenamefont {Hastings},\
  and\ \citenamefont {Troyer}}]{Bauer-2016-1510.03859}%
  \BibitemOpen
  \bibfield  {author} {\bibinfo {author} {\bibfnamefont {B.}~\bibnamefont
  {Bauer}}, \bibinfo {author} {\bibfnamefont {D.}~\bibnamefont {Wecker}},
  \bibinfo {author} {\bibfnamefont {A.~J.}\ \bibnamefont {Millis}}, \bibinfo
  {author} {\bibfnamefont {M.~B.}\ \bibnamefont {Hastings}}, \ and\ \bibinfo
  {author} {\bibfnamefont {M.}~\bibnamefont {Troyer}},\ }\href@noop {}
  {\bibfield  {journal} {\bibinfo  {journal} {arXiv.org}\ ,\ \bibinfo {pages}
  {arXiv:1510.03859 [quant}} (\bibinfo {year} {2016})}\BibitemShut {NoStop}%
\bibitem [{\citenamefont {Rubin}(2016)}]{Rubin-2016-1610.06910}%
  \BibitemOpen
  \bibfield  {author} {\bibinfo {author} {\bibfnamefont {N.~C.}\ \bibnamefont
  {Rubin}},\ }\href@noop {} {\bibfield  {journal} {\bibinfo  {journal}
  {arXiv.org}\ ,\ \bibinfo {pages} {arXiv:1610.06910 [quant}} (\bibinfo {year}
  {2016})}\BibitemShut {NoStop}%
\bibitem [{\citenamefont {Wecker}\ \emph {et~al.}(2014)\citenamefont {Wecker},
  \citenamefont {Bauer}, \citenamefont {Clark}, \citenamefont {Hastings},\ and\
  \citenamefont {Troyer}}]{Wecker:2014aa}%
  \BibitemOpen
  \bibfield  {author} {\bibinfo {author} {\bibfnamefont {D.}~\bibnamefont
  {Wecker}}, \bibinfo {author} {\bibfnamefont {B.}~\bibnamefont {Bauer}},
  \bibinfo {author} {\bibfnamefont {B.~K.}\ \bibnamefont {Clark}}, \bibinfo
  {author} {\bibfnamefont {M.~B.}\ \bibnamefont {Hastings}}, \ and\ \bibinfo
  {author} {\bibfnamefont {M.}~\bibnamefont {Troyer}},\ }\href
  {https://link.aps.org/doi/10.1103/PhysRevA.90.022305} {\bibfield  {journal}
  {\bibinfo  {journal} {Physical Review A}\ }\textbf {\bibinfo {volume} {90}},\
  \bibinfo {pages} {022305} (\bibinfo {year} {2014})}\BibitemShut {NoStop}%
\bibitem [{\citenamefont {Kandala}\ \emph {et~al.}(2017)\citenamefont
  {Kandala}, \citenamefont {Mezzacapo}, \citenamefont {Temme}, \citenamefont
  {Takita}, \citenamefont {Chow},\ and\ \citenamefont
  {Gambetta}}]{Kandala-2017-1704.05018}%
  \BibitemOpen
  \bibfield  {author} {\bibinfo {author} {\bibfnamefont {A.}~\bibnamefont
  {Kandala}}, \bibinfo {author} {\bibfnamefont {A.}~\bibnamefont {Mezzacapo}},
  \bibinfo {author} {\bibfnamefont {K.}~\bibnamefont {Temme}}, \bibinfo
  {author} {\bibfnamefont {M.}~\bibnamefont {Takita}}, \bibinfo {author}
  {\bibfnamefont {J.~M.}\ \bibnamefont {Chow}}, \ and\ \bibinfo {author}
  {\bibfnamefont {J.~M.}\ \bibnamefont {Gambetta}},\ }\href@noop {} {\bibfield
  {journal} {\bibinfo  {journal} {arXiv.org}\ ,\ \bibinfo {pages}
  {arXiv:1704.05018 [quant}} (\bibinfo {year} {2017})}\BibitemShut {NoStop}%
\bibitem [{\citenamefont {Babbush}\ \emph
  {et~al.}(2017{\natexlab{b}})\citenamefont {Babbush}, \citenamefont {Wiebe},
  \citenamefont {McClean}, \citenamefont {McClain}, \citenamefont {Neven},\
  and\ \citenamefont {Chan}}]{Babbush-2017-1706.00023}%
  \BibitemOpen
  \bibfield  {author} {\bibinfo {author} {\bibfnamefont {R.}~\bibnamefont
  {Babbush}}, \bibinfo {author} {\bibfnamefont {N.}~\bibnamefont {Wiebe}},
  \bibinfo {author} {\bibfnamefont {J.}~\bibnamefont {McClean}}, \bibinfo
  {author} {\bibfnamefont {J.}~\bibnamefont {McClain}}, \bibinfo {author}
  {\bibfnamefont {H.}~\bibnamefont {Neven}}, \ and\ \bibinfo {author}
  {\bibfnamefont {G.~K.-L.}\ \bibnamefont {Chan}},\ }\href@noop {} {\bibfield
  {journal} {\bibinfo  {journal} {arXiv.org}\ ,\ \bibinfo {pages}
  {arXiv:1706.00023 [quant}} (\bibinfo {year}
  {2017}{\natexlab{b}})}\BibitemShut {NoStop}%
\bibitem [{\citenamefont {Narimani}\ \emph {et~al.}(2017)\citenamefont
  {Narimani}, \citenamefont {Rezaei},\ and\ \citenamefont
  {Zaribafiyan}}]{NRZ2017}%
  \BibitemOpen
  \bibfield  {author} {\bibinfo {author} {\bibfnamefont {A.}~\bibnamefont
  {Narimani}}, \bibinfo {author} {\bibfnamefont {S.~S.~C.}\ \bibnamefont
  {Rezaei}}, \ and\ \bibinfo {author} {\bibfnamefont {A.}~\bibnamefont
  {Zaribafiyan}},\ }\href@noop {} {\  (\bibinfo {year} {2017})},\ \Eprint
  {http://arxiv.org/abs/arXiv:arXiv:1708.03439v3} {arXiv:arXiv:1708.03439v3}
  \BibitemShut {NoStop}%
\bibitem [{\citenamefont {Dallaire-Demers}\ \emph {et~al.}(2018)\citenamefont
  {Dallaire-Demers}, \citenamefont {Romero}, \citenamefont {Veis},
  \citenamefont {Sim},\ and\ \citenamefont {Aspuru-Guzik}}]{LDCA}%
  \BibitemOpen
  \bibfield  {author} {\bibinfo {author} {\bibfnamefont {P.-L.}\ \bibnamefont
  {Dallaire-Demers}}, \bibinfo {author} {\bibfnamefont {J.}~\bibnamefont
  {Romero}}, \bibinfo {author} {\bibfnamefont {L.}~\bibnamefont {Veis}},
  \bibinfo {author} {\bibfnamefont {S.}~\bibnamefont {Sim}}, \ and\ \bibinfo
  {author} {\bibfnamefont {A.}~\bibnamefont {Aspuru-Guzik}},\ }\href@noop {}
  {\enquote {\bibinfo {title} {Low-depth circuit ansatz for preparing
  correlated fermionic states on a quantum computer},}\ } (\bibinfo {year}
  {2018}),\ \Eprint {http://arxiv.org/abs/arXiv:1801.01053} {arXiv:1801.01053}
  \BibitemShut {NoStop}%
\bibitem [{\citenamefont {Onuchic}\ and\ \citenamefont
  {Wolynes}(2004)}]{Onuchic:2004aa}%
  \BibitemOpen
  \bibfield  {author} {\bibinfo {author} {\bibfnamefont {J.}~\bibnamefont
  {Onuchic}}\ and\ \bibinfo {author} {\bibfnamefont {P.~G.}\ \bibnamefont
  {Wolynes}},\ }\href {\doibase https://doi.org/10.1016/j.sbi.2004.01.009}
  {\bibfield  {journal} {\bibinfo  {journal} {Current Opinion in Structural
  Biology}\ }\textbf {\bibinfo {volume} {14}},\ \bibinfo {pages} {70} (\bibinfo
  {year} {2004})}\BibitemShut {NoStop}%
\bibitem [{\citenamefont {O'Boyle}\ \emph
  {et~al.}(2011{\natexlab{a}})\citenamefont {O'Boyle}, \citenamefont {Banck},
  \citenamefont {James}, \citenamefont {Morley}, \citenamefont
  {Vandermeersch},\ and\ \citenamefont {Hutchison}}]{OBoyle:2011aa}%
  \BibitemOpen
  \bibfield  {author} {\bibinfo {author} {\bibfnamefont {N.~M.}\ \bibnamefont
  {O'Boyle}}, \bibinfo {author} {\bibfnamefont {M.}~\bibnamefont {Banck}},
  \bibinfo {author} {\bibfnamefont {C.~A.}\ \bibnamefont {James}}, \bibinfo
  {author} {\bibfnamefont {C.}~\bibnamefont {Morley}}, \bibinfo {author}
  {\bibfnamefont {T.}~\bibnamefont {Vandermeersch}}, \ and\ \bibinfo {author}
  {\bibfnamefont {G.~R.}\ \bibnamefont {Hutchison}},\ }\href {\doibase
  10.1186/1758-2946-3-33} {\bibfield  {journal} {\bibinfo  {journal} {Journal
  of Cheminformatics}\ }\textbf {\bibinfo {volume} {3}},\ \bibinfo {pages} {33}
  (\bibinfo {year} {2011}{\natexlab{a}})}\BibitemShut {NoStop}%
\bibitem [{\citenamefont {O'Boyle}\ \emph
  {et~al.}(2011{\natexlab{b}})\citenamefont {O'Boyle}, \citenamefont
  {Vandermeersch}, \citenamefont {Flynn}, \citenamefont {Maguire},\ and\
  \citenamefont {Hutchison}}]{OBoyle:2011ab}%
  \BibitemOpen
  \bibfield  {author} {\bibinfo {author} {\bibfnamefont {N.~M.}\ \bibnamefont
  {O'Boyle}}, \bibinfo {author} {\bibfnamefont {T.}~\bibnamefont
  {Vandermeersch}}, \bibinfo {author} {\bibfnamefont {C.~J.}\ \bibnamefont
  {Flynn}}, \bibinfo {author} {\bibfnamefont {A.~R.}\ \bibnamefont {Maguire}},
  \ and\ \bibinfo {author} {\bibfnamefont {G.~R.}\ \bibnamefont {Hutchison}},\
  }\href {\doibase 10.1186/1758-2946-3-8} {\bibfield  {journal} {\bibinfo
  {journal} {Journal of Cheminformatics}\ }\textbf {\bibinfo {volume} {3}},\
  \bibinfo {pages} {8} (\bibinfo {year} {2011}{\natexlab{b}})}\BibitemShut
  {NoStop}%
\bibitem [{\citenamefont {Kobayashi}\ and\ \citenamefont
  {Nakai}(2008)}]{Kobayashi:2008aa}%
  \BibitemOpen
  \bibfield  {author} {\bibinfo {author} {\bibfnamefont {M.}~\bibnamefont
  {Kobayashi}}\ and\ \bibinfo {author} {\bibfnamefont {H.}~\bibnamefont
  {Nakai}},\ }\bibfield  {booktitle} {\emph {\bibinfo {booktitle} {The Journal
  of Chemical Physics}},\ }\href {\doibase 10.1063/1.2956490} {\bibfield
  {journal} {\bibinfo  {journal} {The Journal of Chemical Physics}\ }\textbf
  {\bibinfo {volume} {129}},\ \bibinfo {pages} {044103} (\bibinfo {year}
  {2008})}\BibitemShut {NoStop}%
\bibitem [{\citenamefont {Fedorov}\ and\ \citenamefont
  {Kitaura}(2005)}]{Fedorov:2005aa}%
  \BibitemOpen
  \bibfield  {author} {\bibinfo {author} {\bibfnamefont {D.~G.}\ \bibnamefont
  {Fedorov}}\ and\ \bibinfo {author} {\bibfnamefont {K.}~\bibnamefont
  {Kitaura}},\ }\bibfield  {booktitle} {\emph {\bibinfo {booktitle} {The
  Journal of Chemical Physics}},\ }\href {\doibase 10.1063/1.2007588}
  {\bibfield  {journal} {\bibinfo  {journal} {The Journal of Chemical Physics}\
  }\textbf {\bibinfo {volume} {123}},\ \bibinfo {pages} {134103} (\bibinfo
  {year} {2005})}\BibitemShut {NoStop}%
\bibitem [{\citenamefont {Knizia}\ and\ \citenamefont
  {Chan}(2012)}]{Knizia:2012aa}%
  \BibitemOpen
  \bibfield  {author} {\bibinfo {author} {\bibfnamefont {G.}~\bibnamefont
  {Knizia}}\ and\ \bibinfo {author} {\bibfnamefont {G.~K.-L.}\ \bibnamefont
  {Chan}},\ }\href {https://link.aps.org/doi/10.1103/PhysRevLett.109.186404}
  {\bibfield  {journal} {\bibinfo  {journal} {Physical Review Letters}\
  }\textbf {\bibinfo {volume} {109}},\ \bibinfo {pages} {186404} (\bibinfo
  {year} {2012})}\BibitemShut {NoStop}%
\bibitem [{\citenamefont {Wouters}\ \emph {et~al.}(2016)\citenamefont
  {Wouters}, \citenamefont {Jim{\'e}nez-Hoyos}, \citenamefont {Sun},\ and\
  \citenamefont {Chan}}]{Wouters:2016aa}%
  \BibitemOpen
  \bibfield  {author} {\bibinfo {author} {\bibfnamefont {S.}~\bibnamefont
  {Wouters}}, \bibinfo {author} {\bibfnamefont {C.~A.}\ \bibnamefont
  {Jim{\'e}nez-Hoyos}}, \bibinfo {author} {\bibfnamefont {Q.}~\bibnamefont
  {Sun}}, \ and\ \bibinfo {author} {\bibfnamefont {G.~K.~L.}\ \bibnamefont
  {Chan}},\ }\bibfield  {booktitle} {\emph {\bibinfo {booktitle} {Journal of
  Chemical Theory and Computation}},\ }\href {\doibase
  10.1021/acs.jctc.6b00316} {\bibfield  {journal} {\bibinfo  {journal} {Journal
  of Chemical Theory and Computation}\ }\textbf {\bibinfo {volume} {12}},\
  \bibinfo {pages} {2706} (\bibinfo {year} {2016})}\BibitemShut {NoStop}%
\bibitem [{\citenamefont {Fedorov}\ and\ \citenamefont
  {Kitaura}(2009)}]{Fedorov-2009}%
  \BibitemOpen
  \bibfield  {author} {\bibinfo {author} {\bibfnamefont {D.}~\bibnamefont
  {Fedorov}}\ and\ \bibinfo {author} {\bibfnamefont {K.}~\bibnamefont
  {Kitaura}},\ }\href@noop {} {\emph {\bibinfo {title} {The Fragment Molecular
  Orbital Method: Practical Applications to Large Molecular Systems}}}\
  (\bibinfo  {publisher} {CRC Press},\ \bibinfo {address} {6000 Broken Sound
  Parkway NW, Suite 300, Boca Raton, FL 33487-2742},\ \bibinfo {year}
  {2009})\BibitemShut {NoStop}%
\bibitem [{\citenamefont {Schmidt}\ \emph {et~al.}(1993)\citenamefont
  {Schmidt}, \citenamefont {Baldridge}, \citenamefont {Boatz}, \citenamefont
  {Elbert}, \citenamefont {Gordon}, \citenamefont {Jensen}, \citenamefont
  {Koseki}, \citenamefont {Matsunaga}, \citenamefont {Nguyen}, \citenamefont
  {Su}, \citenamefont {Windus}, \citenamefont {Dupuis},\ and\ \citenamefont
  {Montgomery}}]{Schmidt:1993aa}%
  \BibitemOpen
  \bibfield  {author} {\bibinfo {author} {\bibfnamefont {M.~W.}\ \bibnamefont
  {Schmidt}}, \bibinfo {author} {\bibfnamefont {K.~K.}\ \bibnamefont
  {Baldridge}}, \bibinfo {author} {\bibfnamefont {J.~A.}\ \bibnamefont
  {Boatz}}, \bibinfo {author} {\bibfnamefont {S.~T.}\ \bibnamefont {Elbert}},
  \bibinfo {author} {\bibfnamefont {M.~S.}\ \bibnamefont {Gordon}}, \bibinfo
  {author} {\bibfnamefont {J.~H.}\ \bibnamefont {Jensen}}, \bibinfo {author}
  {\bibfnamefont {S.}~\bibnamefont {Koseki}}, \bibinfo {author} {\bibfnamefont
  {N.}~\bibnamefont {Matsunaga}}, \bibinfo {author} {\bibfnamefont {K.~A.}\
  \bibnamefont {Nguyen}}, \bibinfo {author} {\bibfnamefont {S.}~\bibnamefont
  {Su}}, \bibinfo {author} {\bibfnamefont {T.~L.}\ \bibnamefont {Windus}},
  \bibinfo {author} {\bibfnamefont {M.}~\bibnamefont {Dupuis}}, \ and\ \bibinfo
  {author} {\bibfnamefont {J.~A.}\ \bibnamefont {Montgomery}},\ }\href
  {\doibase 10.1002/jcc.540141112} {\bibfield  {journal} {\bibinfo  {journal}
  {Journal of Computational Chemistry}\ }\textbf {\bibinfo {volume} {14}},\
  \bibinfo {pages} {1347} (\bibinfo {year} {1993})}\BibitemShut {NoStop}%
\bibitem [{\citenamefont {Wouters}()}]{Wouters-QC-DMET}%
  \BibitemOpen
  \bibfield  {author} {\bibinfo {author} {\bibfnamefont {S.}~\bibnamefont
  {Wouters}},\ }\href@noop {} {\ }\BibitemShut {NoStop}%
\bibitem [{\citenamefont {Sun}\ \emph {et~al.}(2018)\citenamefont {Sun},
  \citenamefont {Berkelbach}, \citenamefont {Blunt}, \citenamefont {Booth},
  \citenamefont {Guo}, \citenamefont {Li}, \citenamefont {Liu}, \citenamefont
  {McClain}, \citenamefont {Sayfutyarova}, \citenamefont {Sharma},
  \citenamefont {Wouters},\ and\ \citenamefont {Chan}}]{Sun:2018aa}%
  \BibitemOpen
  \bibfield  {author} {\bibinfo {author} {\bibfnamefont {Q.}~\bibnamefont
  {Sun}}, \bibinfo {author} {\bibfnamefont {T.~C.}\ \bibnamefont {Berkelbach}},
  \bibinfo {author} {\bibfnamefont {N.~S.}\ \bibnamefont {Blunt}}, \bibinfo
  {author} {\bibfnamefont {G.~H.}\ \bibnamefont {Booth}}, \bibinfo {author}
  {\bibfnamefont {S.}~\bibnamefont {Guo}}, \bibinfo {author} {\bibfnamefont
  {Z.}~\bibnamefont {Li}}, \bibinfo {author} {\bibfnamefont {J.}~\bibnamefont
  {Liu}}, \bibinfo {author} {\bibfnamefont {J.~D.}\ \bibnamefont {McClain}},
  \bibinfo {author} {\bibfnamefont {E.~R.}\ \bibnamefont {Sayfutyarova}},
  \bibinfo {author} {\bibfnamefont {S.}~\bibnamefont {Sharma}}, \bibinfo
  {author} {\bibfnamefont {S.}~\bibnamefont {Wouters}}, \ and\ \bibinfo
  {author} {\bibfnamefont {G.~K.-L.}\ \bibnamefont {Chan}},\ }\href {\doibase
  10.1002/wcms.1340} {\bibfield  {journal} {\bibinfo  {journal} {Wiley
  Interdisciplinary Reviews: Computational Molecular Science}\ }\textbf
  {\bibinfo {volume} {8}},\ \bibinfo {pages} {n/a} (\bibinfo {year}
  {2018})}\BibitemShut {NoStop}%
\bibitem [{\citenamefont {Hehre}\ \emph {et~al.}(1972)\citenamefont {Hehre},
  \citenamefont {Ditchfield},\ and\ \citenamefont {Pople}}]{Hehre:1972aa}%
  \BibitemOpen
  \bibfield  {author} {\bibinfo {author} {\bibfnamefont {W.~J.}\ \bibnamefont
  {Hehre}}, \bibinfo {author} {\bibfnamefont {R.}~\bibnamefont {Ditchfield}}, \
  and\ \bibinfo {author} {\bibfnamefont {J.~A.}\ \bibnamefont {Pople}},\
  }\bibfield  {booktitle} {\emph {\bibinfo {booktitle} {The Journal of Chemical
  Physics}},\ }\href {\doibase 10.1063/1.1677527} {\bibfield  {journal}
  {\bibinfo  {journal} {The Journal of Chemical Physics}\ }\textbf {\bibinfo
  {volume} {56}},\ \bibinfo {pages} {2257} (\bibinfo {year}
  {1972})}\BibitemShut {NoStop}%
\bibitem [{\citenamefont {Dunning}(1989)}]{Dunning:1989aa}%
  \BibitemOpen
  \bibfield  {author} {\bibinfo {author} {\bibfnamefont {T.~H.}\ \bibnamefont
  {Dunning}},\ }\bibfield  {booktitle} {\emph {\bibinfo {booktitle} {The
  Journal of Chemical Physics}},\ }\href {\doibase 10.1063/1.456153} {\bibfield
   {journal} {\bibinfo  {journal} {The Journal of Chemical Physics}\ }\textbf
  {\bibinfo {volume} {90}},\ \bibinfo {pages} {1007} (\bibinfo {year}
  {1989})}\BibitemShut {NoStop}%
\end{thebibliography}%
\end{document}